\documentclass[11pt]{article} 
\usepackage{natbib}
 \bibpunct{(}{)}{;}{a}{,}{,}
\usepackage{amsmath}
\usepackage{amsthm}
\usepackage{amsfonts}
\usepackage{amssymb}
\usepackage{tabularx}
\usepackage{epsfig}
\usepackage{dsfont}
\usepackage{natbib}
\usepackage{geometry}
\newtheorem{thm}{Theorem}[section]

\let\pf=\proof

\newcommand{\bv}{\begin{array}}
\newcommand{\ev}{\end{array}}
\newcommand{\bit}{\begin{itemize}}
\newcommand{\eit}{\end{itemize}}
\newcommand{\ben}{\begin{enumerate}}
\newcommand{\een}{\end{enumerate}}
\newcommand{\beq}{\begin{equation}}
\newcommand{\eeq}{\end{equation}}
\newcommand{\bvq}{\begin{eqnarray}}
\newcommand{\evq}{\end{eqnarray}}

\begin{document}

\title{Valid inference from non-ignorable network sampling mechanisms}


\author{Sim\'on Lunag\'omez \\
	Department of Statistics, University College London, London WC1E 6BT, UK.  \\
	s.lunagomez@ucl.ac.uk\\
	\and 
	Edoardo M Airoldi \\
	Department of Statistics, Harvard University, Cambridge, MA 02138, USA. \\
	airoldi@fas.harvard.edu\\
	}

 
 \maketitle


\begin{abstract}
Consider a population of individuals and a  network that encodes social connections among them.   We are interested in making inference on finite population and super-population estimands that are a function of both individuals' responses and of the network, from a sample. Neither the sampling frame nor the network are available. However, the sampling mechanism implicitly leverages the network to recruit individuals, thus partially revealing social interactions among the individuals in the sample, as well as their responses.   This is a common setting that arises, for instance, in epidemiology and healthcare, where samples from hard-to-reach populations are collected using link-tracing mechanisms, including respondent-driven sampling. Contrary to random sampling, the probability models of these network sampling mechanisms carry information about the estimands of interest, such as the incidence of certain diseases in the target population.   In this paper, we study statistical properties of popular network sampling mechanisms. We formulate the estimation problem by extending Rubin's inferential framework to explicitly   account for social network structure. We then identify key modeling elements that lead to inferences with good frequentist properties when dealing with data collected through non-ignorable network sampling mechanisms.   We demonstrate these methods on a study of the  incidence of HIV in Brazil.
%
%
%
\end{abstract}



\section{Introduction}
\label{sec:introduction}







Consider a population of individuals that is susceptible to a disease, e.g. AIDS, it is of interest to estimate the prevalence of the disease among high-risk individuals, for example men that have sex with men or intravenous drug abusers. It has been observed that the most effective methods to answer these questions, so far, rely on sampling mechanisms that take advantage of the social network structure, e.g., respondent-driven sampling (RDS) (See \cite{Hecka} and \cite{VolzHecka}).

Let $Y(1),Y(2)$,\dots $Y(N)$ be a vector of responses (e.g. the results of the HIV test), $I$ the sampling mechanism (e.g. RDS) and $\mathcal{G}$, the social network. We are interested in estimating a population quantity $Q=Q(Y,\mathcal{G})$ which is a function of the response vector and the social network. In order to perform likelihood-based or Bayesian inference on $Q$, it is necessary to specify the likelihood correctly. Part of this specification involves determining if the uncertainty regarding $I$ is relevant or not \citep{Rubin2}. 

When the sampling mechanisms is ignorable, not including explicitly a probability model for it in the likelihood  does not have an impact on the inference (either likelihood-based or Bayesian) of a population quantity. The only information that is needed for good inference are the indicators of the individuals included in the sample. \cite{Rubin2} and \cite{HeitjRubin} have developed a rigorous approach for tackling the question of when it is valid to ignore the functional form of the sampling mechanism for performing inferences. A key notion for this approach is the one of \emph{ignorability} \citep{Rubin1}, which establishes when the probability distribution of the sampling design is relevant for modeling the distribution of random quantities corresponding to individuals not included in the sample. Under Rubin's framework ignorability is equivalent to saying that the posterior of the population quantity can be computed without conditioning on the functional form of the sampling design. A sampling design is called \emph{non-ignorable} if its functional form has to be expressed explicitly in the model in order to perform likelihood-based or Bayesian inference.

\begin{figure}[t!]
\centering
\epsfig{file=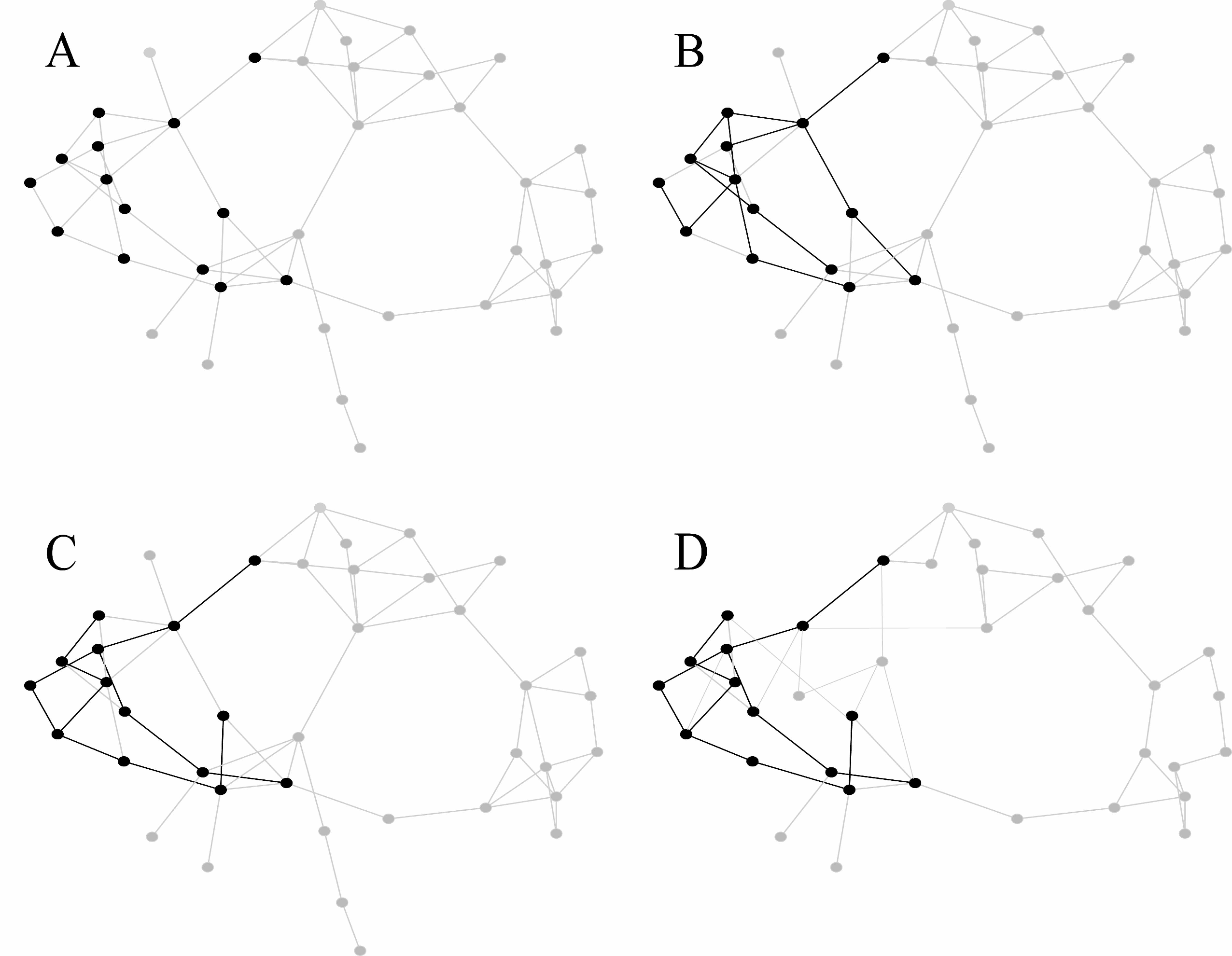,width=4.5in}
\caption{An illustration of the sources of non-ignorability in link-tracing designs. 
         {\it Panel A:} A population graph and a random sample. 
         {\it Panel B:} A realization of a link-tracing design on that produces the sample in panel A.
         {\it Panel C:} A different realization of a link-tracing design on that produces the same sample in panel A.
         {\it Panel D:} The same realization of a link-tracing design in panel C, but on a different population graph; it has a different likelihood.}
\label{Fig:IllustrateGraph}
\end{figure}

We consider a situation where non-ignorability arises because the sampling design is driven by a network, which is progressively discovered through sampling. One way to understand this is that the likelihood will depend on quantities indexed to the individuals not sampled. This could happen in at least two ways.
First, the probability distribution of the sampling design depends on features corresponding the portion of the graph that was not sampled; this is illustrated by comparing panels C and D in Figure \ref{Fig:IllustrateGraph}. An obvious implication of this is that changes in the underlying network will affect the likelihood. An equally important, but greatly overlooked aspect of this is that we could have different realizations of the sampling mechanism, leading to the same set of sampled individuals, but conveying different information about the network structure, therefore producing different inferences; this point is illustrated by comparing panels B and C in Figure \ref{Fig:IllustrateGraph}.
Second, the network induces a dependence structure on the responses, in such case the responses of not sampled individuals need to be taken into account for computing the likelihood.
%

Another source of non-ignorability may arise at the response level, \emph{i.e.}, given that node $i$ in the network is included in the sample, the value of $y(i)$ (the realization of $Y(i)$) of the response could be missing with a probability that depends on the value of $y(i)$. While non ignorable response is not testable, non ignorable sampling designs are, since the analysts is implementing the data collection protocol. In practice both sources of ignorability may be confounded, and often are, but still our theory is relevant to let only the non-ignorable response be the thing to worry in the analysis.

From a statistical perspective, the issues of inferring a population quantity using a non-ignorable sampling design on a social network include: Modeling the unobserved part of the social network in probabilistic fashion, understanding the sampling mechanism as a probability model and including it in the likelihood, and finally, modeling the dependence structure of the responses given the network. Not taking into account any of these issues leads to misleading inferences and drastic understatements regarding the uncertainty associated to those inferences. This has been preliminary examined in \cite{Orbe}.
A sample design based on a social network structure that is widely used for inferring a population quantity is Respondent-Driven Sampling (RDS). This design was proposed by \cite{Hecka}. Later \cite{VolzHecka} proposed an estimation procedure tailored for RDS based on the assumption that the relationship between the probability of inclusion of a given individual and the degree of the corresponding node is linear. \cite{Gile} improved this methodology by estimating the relationship between inclusion probability and degree distribution via an iterative procedure. In the same spirit \cite{AronCraw} estimate the inclusion probability when the relationship with respect to the degrees is known up to scale. None of this approaches is model-based, therefore, they are all vulnerable to issues involving, either the non-ignorability of the sampling mechanism, or the dependence among the responses. What is more relevant to the discussion is that these approaches assume that RDS is an ignorable design; we prove this is not true (Section \ref{Sec:RDS}). A very interesting work that involves the concept of ignorability is \cite{HandcGile}. Their focus is on estimating the parameters of the social network model, not in estimating a population quantity while allowing uncertainty for the network structure. None of the previous literature deals with the problem of non-ignorable designs on networks. A recent paper by \cite{GileJohnSalg} deals with diagnostics for checking the assumptions of RDS.

In this paper we propose methodology that achieves multiple goals.
(1) Allows us to perform inferences in cases where the sampling mechanism is non-ignorable.
(2) Takes into account all relevant sources of uncertainty, \emph{i.e.}, uncertainty regarding the underlying social network, the sampling mechanism  and the parameters of the model.
(3) Models the dependence structure of the response explicitly by using the social network structure.
(4) Is modular, in the sense that different priors and likelihoods can be used for the components of the model.
(5) Allows inference for both finite population and infinite super population estimands.
We developed a general framework that accomplishes these five objectives. The underlying network structure is understood as a statistical network model. The dependence structure of the responses given the social network is modeled via a Markov Random Field (MRF). We write likelihoods for the sampling mechanisms given the graph. Inferences on the model are performed using a Bayesian approach \citep{Rober}. One of the challenges we had to deal with was that only part of the network is observed, and the number of unobserved nodes and edges is unknown, this involves a problem of variable dimension. We solved this difficulty by employing Bayesian model averaging (see \cite{RafteMadigVolin} and \cite{Rober}, Section 7.4). Once the dimension is fixed, MCMC techniques for MRF's are developed, this was particularly hard because the joint distribution for an MRF cannot be written in close form (all is needed is to write the quotients for the Metropolis ratio). An important contribution of this paper is that it departs from the generality of Rubin's framework in the sense that instead of considering a joint distribution for all the elements in the model with no structure at all, we impose a series of conditional independence statements among them that correspond to reasonable general assumptions about sampling on social networks problems. These conditional independence statements and the strategy we propose for computing the posterior (Section \ref{sec-inference}), allow our approach to be modular: \emph{i.e.}, our methodology enables the statistician to propose different distributions for the random graph model and the MRF, given the sampling mechanism.

A  surprising feature of our framework is that it allows us to circumvent the assumptions established in \cite{SalgaHecka} and \cite{Hecka2} in order to guarantee consistency of the Volz-Heckathorn estimator for RDS sampling. We elaborate on this point in the discussion.

The outline of the paper goes as follows: In Section \ref{Sec:Theory} we phrase the problem in terms of the general framework established by \cite{Rubin1} and set the basis for talking rigorously about the role of knowing the distribution of the sampling mechanism for performing inferences on the population quantity. In Section \ref{Sec:Model} we present a general framework for performing Bayesian inferences about a population quantity in a social network context. We also provide specific choices for the distributions required by the model and a description of the Markov Chain Monte Carlo (MCMC) scheme for this model (Section \ref{Sec:Results}). Simulations illustrating the advantages of the proposed methodology are shown in Section \ref{Sec:Results}. A case study is presented in Section \ref{Sec:RealData}. Section \ref{Sec:Discuss} is the discussion.

\section{Theory and Definitions}\label{Sec:Theory}

\subsection{The Concept of Ignorability in the Context of Social Networks}

  Let  $W$ denote the full data, with distribution $p(W\mid \tau)$, and let $I$ denote a sampling mechanism, \emph{i.e.}, the dynamic process of data collection. We adopt the convention (proposed by \citep{Rubin1}) of using the letter $I$ to also represent the indicator for the data included in the sample. Let  $W_{\text{INC}}$ denote the observed data \emph{i.e.}, the elements of $W$ for which $I=1$. The joint distribution of $(W,I)$ is often written as
 \begin{equation}\label{Eq:RubinFactor}
 p(W,I,\tau,\eta)=p(W\mid \tau)p(I \mid W, \eta),
\end{equation}  
 where $\eta$ represents the tuning parameters of the sampling mechanism. A sampling mechanism $I$ is called \emph{ignorable} (\cite{Rubin1}) if
\begin{equation}\label{Eq:MisRand}
p(I \mid W,\eta)=p(I \mid W_{\text{INC}},\eta)
\end{equation}
and the parameters for the sampling mechanism ($\eta$) and the full data ($\tau$) are distinct. If a sample mechanism $I$ does not fulfil these conditions, it is called \emph{non-ignorable}.   The condition stated in Equation \ref{Eq:MisRand} is called \emph{missing at random}.
 If a sampling mechanism is ignorable, then the term corresponding  to the distribution of $I$ can be omitted from the likelihood. By this it is meant that the likelihood for $\tau$ is obtained by integrating out the missing data from $p(W\mid \tau)$. In contrast, when the sampling mechanism is non-ignorable, the likelihood for $\tau$ is obtained by integrating out the missing data from $p(W\mid \tau)p(I \mid W, \eta)$.

For the purpose of this paper, it is reasonable to set $W=(Y,\mathcal{G})$, where $\mathcal{G}=(\mathcal{V},\mathcal{E})$ is the graph or social network, with  vertex set $\mathcal{V}$ and edge set $\mathcal{E}$, and $Y$ is the vector of univariate responses associated to each node of $\mathcal{G}$. This means that a sampling mechanism $I$ is non-ignorable if its distribution is not constant with respect to unobserved entries of $Y$ or to features of the unobserved portion of the network. Although we are merging $Y$ and $\mathcal{G}$ into $W$, note that conceptually they are quite different entities: $Y$ is the variable of interest for the investigator, while $\mathcal{G}$ provides information regarding the connections among individuals, which can be used to weight the $Y$'s or to suggest a dependence structure \emph{i.e.}, the observed part of $\mathcal{G}$ works as a set of post-sampling covariates. We denote by $\mathcal{G}_{\text{INC}}$ and $\mathcal{G}_{\text{EXC}}$, respectively, the observed and unobserved parts of the network. $Y_{\text{INC}}$ denotes the responses of the individuals that were included in the sample, $Y_{\text{EXC}}$ denotes the responses of the individuals that were not included in the sample. 

The notion of \emph{amenability} discussed in \citep{HandcGile} is a particular case of the definition of ignorability proposed in \citep{Rubin1}. Such notion is useful when the object of the inference is the parameter vector that serves to specify the distribution of $\mathcal{G}$. In this paper, we will adopt Rubin's definition of ignorability, since it is general enough to encompass the cases where the object of the inference is either:
\begin{enumerate}
\item A quantity of the form $Q=Q(\tau)$, where $\tau$ serves to specify the distribution of the random graph and the joint distribution of the responses.
\item A quantity of the form $Q=Q(\mathcal{G},Y)$, \emph{e.g.} an average of the response vector weighted by the node degrees. 
\end{enumerate}
An estimand of the form $Q=Q(\tau)$ will be used to pose inferential questions at a super-population level (these will be statements in terms of parameters). To pose inferential questions at a population level, we will use an estimand of the form $Q=Q(\mathcal{G},Y)$ (these will be statements about the missing data distribution). Both estimands will be discussed in detail in Section \ref{sec-estimands}.

\subsection{Respondent-Driven Sampling}\label{Sec:RDS}

Respondent-Driven Sampling (RDS) is a sampling procedure that takes advantage of the social network structure. It was proposed by \cite{Hecka}. RDS consists of choosing $s$ starting points (individuals identified as nodes of the network) according to a pre-specified mechanism that can be either deterministic or random, and then propagating the sample by following a set of policies defined in terms of the social network.  Let $N$ denote $|\mathcal{V}|$, the number nodes for $\mathcal{G}$, $n$ denote the sample size, and $m$ represent the maximum number of referrals per node. Let $D$ denote the vector that has as $i-$th component the degree of the $i-$th vertex, and let $D_\text{INC}$ denote the vector of observed degrees. We describe the RDS algorithm in terms of the construction of $(\mathcal{G}_{\text{INC}}, Y_{\text{INC}}, D_{\text{INC}})$:  
\begin{enumerate}
\item Sample $s$ nodes uniformly from $\mathcal{V}$. This is known as the $0$-\emph{th wave}. The selected nodes constitute $\mathcal{G}_{\text{INC}[0]}$.
\item For each node $i$ in Step (a), record the response $Y(i)$ in $Y_{\text{INC}[0]}$ and the corresponding degree $D(i)$ in $D_{\text{INC}[0]}$. 
\item For each node in the $(k-1)$-\emph{th wave}, sample uniformly $m$ nodes among its neighbours relative to $\mathcal{G}$, such that they have not been included in $\mathcal{G}_{\text{INC}}$ already. This is known as the $k$-\emph{th wave}. 
\item For each node $i$ sampled in Step (c), record the response $Y(i)$ in $Y_{\text{INC}[k]}$, the corresponding degree $D(i)$ in $D_{\text{INC}[k]}$, and the edge that connected $i$ to $\mathcal{G}_{\text{INC}[k-1]}$ to construct $\mathcal{G}_{\text{INC}[k]}$.
\item Repeat Steps (c) and (d) until the pre-specified sample size $n$ has been attained. Interrupt the current wave if necessary.
\end{enumerate}
Let $\eta$ denote the vector of tuning parameters $(m,s,n)$. We use steps (a)-(e) to describe RDS as an algorithm with inputs $(\mathcal{G},Y, \eta)$ and outputs $(\mathcal{G}_{\text{INC}}, Y_{\text{INC}}, D_{\text{INC}})$. In our notation, $\mathcal{G}_{\text{INC}}$ denotes not only the observed subgraph, but it also encodes the order in which edges were added to it. Observe that Step (a) was set this way in order to ease the exposition. The distribution of the starting points can be modified (in particular, it can be set as a point mass). The description we provided for step (c) is useful when implementing a simulation, however, in practice, it is the individual recruited in the study who selects $m$ out of his\slash her contacts; neither the statistician nor the practitioner implementing the sampling know, in advance, the list of contacts of any of the individuals involved in the study. Finally, note that, the notion of wave encodes the discrete time involved in the sampling process. 

Remember that, in this paper, we follow the convention established by  \cite{Rubin1} of incorporating the uncertainty of the sampling mechanism via its conditional distribution with respect of the full data (Equation \ref{Eq:RubinFactor}). In principle, one should write the distribution of RDS as $p(I \mid \mathcal{G},Y,\eta)$. This intuition seems to be validated by the fact that RDS can be seen as a stochastic algorithm with $(\mathcal{G},Y,\eta)$ as inputs. However, from the policies $(a)-(e)$, it follows that the distribution of RDS is constant with respect to the value of $Y$. Therefore, from now on, the distribution of RDS (and any other link-tracing design) will be written as $p(I \mid \mathcal{G},\eta)$.

Let $I$ denote RDS, which is understood as a probabilistic process for gathering data that propagates through a network $\mathcal{G}$. RDS is fully specified by $\eta=(m,s,n)$. The probability distribution for $I$ given  $\mathcal{G}$ ($s$ was set equal to 1, to ease the exposition) can be written as:
\begin{equation}\label{Eq:DistCondRDS}
p(I \mid \mathcal{G},\eta)=\frac{1}{\binom{\tilde{d}_0}{m}} \left( \prod_{j_1=1}^{w_0} \frac{1}{\binom{\tilde{d}_{j_1}}{m}}  \left[     \prod_{j_2,j_1=1}^{w_{j_1}} \frac{1}{\binom{\tilde{d}_{j_2,j_1}}{m}}   \cdots   \left[    \prod_{j_k,\dots,j_1=1}^{w_{j_{k-1},\dots,j_1}}   \frac{1}{\binom{\tilde{d}_{j_k,\dots,j_1}}{m}}    \right]   \cdots  \right]   \right).
\end{equation}
Here $\tilde{d}_{(\cdot)}$ stands for the number of neighbours (with respect to $\mathcal{G}$) of a given vertex that have not been sampled before it, we refer to it as the \emph{adjusted degree}; $w_{(\cdot)}$ denotes the number of recruited individuals during the previous wave by a given vertex (clearly, $w_{(\cdot)} \leq m$);  while $k$ denotes the number of waves needed to recruit $n$ individuals (such number is a function of the RDS policies and the restrictions imposed by the graph topology).


\begin{thm}\label{p:ignore}
Respondent-Driven Sampling is non-ignorable. 
\end{thm}
\begin{proof}
\pf
For RDS to be ignorable, it is necessary that all the quantities involved in Equation \ref{Eq:DistCondRDS} can be computed from $(\mathcal{G}_{\text{INC}}, Y_{\text{INC}}, D_{\text{INC}})$. To obtain the exact values of $\tilde{d}_{(i)}$, for $i\in\left\{ 1,2,\dots,n \right\}$, one needs to know which neighbours of $\mathcal{V}(i)$ have been added to the vertex set of  $\mathcal{G}_{\text{INC}}$ before it, but the only pieces of information available to answer that question are: the total number of neighbours of $\mathcal{V}(i)$ and the node that referred it (if $\mathcal{V}(i)$ is not a seed). If the degree of $\mathcal{V}(i)$ is greater than $m+1$, then the information available is insufficient to recover  $\tilde{d}_{(i)}$. It follows that RDS is non-ignorable.
\end{proof}

\section{Statistical Methodology}\label{Sec:Model}
\subsection{Modeling framework}
\label{sec-model}
We assume a data generative model of the form:
\begin{equation}\label{Eq:GFramework}
p(Y,I,\mathcal{G},\alpha,\gamma)=p(\alpha)p(\mathcal{G} \mid \alpha) p(I \mid \mathcal{G},\eta) p(\gamma)p(Y \mid \mathcal{G},\gamma). 
\end{equation}
This means that we understand $\mathcal{G}$ as a realization of a random graph (statistical network) model with parameter vector $\alpha$. As explained before, the distribution of $I$ is conditional on a given realization of $\mathcal{G}$, it is not necessary to know how the graph was generated to sample $I$ (an assumption of link-tracing designs). Therefore $\alpha$ does not appear on the conditional for $I$. A key assumption of our approach is that $\mathcal{G}$ induces a dependence structure on $Y$. To fully specify such dependence structure, additional parameters may be needed; those parameters are denoted by $\gamma$, that explains the term $p(Y \mid  \mathcal{G}, \gamma)$. $p(\alpha)$ and $p(\gamma)$ are the priors for $\alpha$ and $\gamma$, respectively. 

To simulate from the generative model proposed in Equation \ref{Eq:GFramework}, first one needs to generate the pair $(\mathcal{G},\alpha)$, then, a realization of $\gamma$ to be able to sample $Y\mid(\mathcal{G},\gamma)$. Once one has $(\mathcal{G},Y)$, the sample mechanism $I$ is applied. In the case of RDS, this produces $(\mathcal{G}_{\text{INC}}, Y_{\text{INC}}, D_{\text{INC}})$ as the data one would be able to observe.

Observe that we include the factor $p(I \mid \mathcal{G},\eta)$ to deal with non-ignorability issues. Still, to achieve this, we need to model the missing data, which, in this case is given by the underlying graph: that is the purpose of adding the factor $p(\mathcal{G} \mid \alpha) p(\alpha)$. For this paper, we assume that the responses are binary, more specifically, that each $y(i)$ takes values in $\left\{ 0,1 \right\}$. In addition, the factor $p(Y \mid \mathcal{G},\gamma)$  will always be assumed to represent a Markov Random Field (MRF).

Note that Expression \ref{Eq:GFramework} is compatible with the factorisation described in Expression \ref{Eq:RubinFactor}. Here $W=(Y,\mathcal{G})$ and $\tau=(\alpha,\gamma)$. For the problems discussed in this paper, the sampling mechanism will be driven by tuning parameters $(\eta)$ that are known to the researcher, and therefore, they are not part of the inference. For the sake of simplicity, we do not include covariates, but there is no impediment for incorporating them if needed.

\subsection{Estimands}
\label{sec-estimands}
The quantities to be estimated are: the proportion of positive responses at the population and super-population levels. In the context of this paper, each response corresponds to a measurement performed on one of the $N$ individuals associated to the nodes of a network $\mathcal{G}$. We first present the estimand for the super-population inference:
\begin{equation}\label{Eq:TheQDef}
Q_\bigstar(\alpha,\gamma)=\mathbb{E}\left( \bar{Y}  \mid \alpha, \gamma \right)=\mathbb{E}\left[ \mathbb{E}\left(  \bar{Y}\mid  \mathcal{G},\gamma \right)  \mid \alpha\right],
\end{equation}
here, the expectation inside the parenthesis is with respect to $p(Y\mid \mathcal{G},\gamma)$ and the one outside the parenthesis is with respect to $p(\mathcal{G}\mid \alpha)$. The estimand associated to the population-level inference is of the form: 
\begin{equation}\label{Eq:TheQPred}
Q_\circ(\mathcal{G}_{\text{INC}}, Y_{\text{INC}})=\mathbb{E}\left[ \mathbb{E}\left(  \frac{1}{N} \left( \sum_{i=1}^n Y_{\text{INC}}(i) + \sum_{k=1}^{N-n}Y_{\text{EXC}}(k)  \right)   \mid  \mathcal{G}_{\text{EXC}} \right)  \mid \mathcal{G}_{\text{INC}}, Y_{\text{INC}}\right],
\end{equation}
where the expectation inside the parenthesis is with respect to $p(Y_{\text{EXC}}\mid \mathcal{G}_{\text{EXC}},\mathcal{G}_{\text{INC}},Y_{\text{INC}})$, while the expectation outside the parenthesis is with respect to $p(\mathcal{G}_{\text{EXC}} \mid \mathcal{G}_{\text{INC}},Y_{\text{INC}})$.

\subsection{Posterior inference and estimation}
\label{sec-inference}

Because of the MRF assumption, to compute the likelihood of $Y_{\text{INC}}$, it is necessary to augment versions of $(Y_{\text{EXC}},\mathcal{G}_{\text{EXC}})$ to the observed data. Augmenting $\mathcal{G}_{\text{EXC}}$ is also a key step for dealing with the non-ignorability of the sampling mechanism (as explained in Sections \ref{Sec:RDS} and \ref{sec-model}). Since we do not know how many nodes and edges were unobserved due to $I$, the model becomes one of variable dimension. For the sampling mechanisms we considered for this paper, the way $\mathcal{G}_{\text{INC}}$ is augmented will have an impact on the factor of the likelihood that includes $I$, \emph{i.e.}, they are non-ignorable.

To perform inference on the model we just proposed we will pursue a Bayesian model averaging (see \citep{RafteMadigVolin} and \citep{Rober}, Section 7.4) strategy:
\begin{align}\label{Eq:BMA4RDS}
p(Q_\bigstar \mid Y_{\text{INC}}, \mathcal{G}_{\text{INC}}, D_{\text{INC}}) & =\sum_w p(\mathcal{G}_{{\text{EXC}},w}, \alpha_w\mid  \mathcal{G}_{\text{INC}}, D_{\text{INC}}) \int_{\Gamma(\mathcal{G}_{\text{I}},\mathcal{G}_{{\text{EXC}},w})} p_w(Q_\bigstar \mid \alpha_w,\gamma_w) \nonumber \\
& \times p(\gamma_w \mid Y_{\text{INC}}, \mathcal{G}_{\text{INC}},\mathcal{G}_{{\text{EXC}},w}) d\gamma_w.
\end{align}
In this setting, the mixing distribution $p(\mathcal{G}_{\text{EXC},w}, \alpha_w\mid  \mathcal{G}_{\text{INC}}, D_{\text{INC}})$ provides information about the random graph model and an imputation of the underlying network,
the posterior  $p(\gamma_w \mid Y_{\text{INC}}, \mathcal{G}_{\text{INC}},\mathcal{G}_{\text{EXC},w})$ encodes what has been learned about the parameter vector driving the dependence structure of the $Y$'s, and
$p_w(Q_\bigstar \mid \alpha_w,\gamma_w)$ produces a Monte Carlo version of $Q_\bigstar$ given $(\alpha_w,\gamma_w)$ (See Expression \ref{Eq:TheQDef}). Here, $w$ indexes the network data imputations implied by $(\mathcal{G}_{{\text{EXC}},w}, \alpha_w)$, where each imputation fixes the dimension of the problem. Equation \ref{Eq:BMA4RDS} provides a reasonable strategy for computing the posterior for $Q_\bigstar$ as long as reliable inferences on $(\mathcal{G}_{\text{EXC}},\alpha)$ can be performed without resorting to information contained in $Y_{\text{INC}}$. Here $\Gamma(\cdot,\cdot,\cdot)$ denotes the set of values $\gamma$ that imply a valid probability model for $Y$, given an imputed network. 

So far, we have referred to $\mathcal{G}_{\text{EXC}}$ as the `unobserved part of the network', however, to perform the computations implied by Equation \ref{Eq:BMA4RDS}, more precision is required. Given $\mathcal{G}_{\text{INC}}$ and $N$ (the maximum allowed size for an imputed network), $\mathcal{G}_{\text{EXC}}$ is given by a set of nodes $\left\{  n+1,n+2,\dots, N_{*} \right\}$, where $N_{*} \leq N$ and a set of edges for an adjacency matrix with $N_{*}$ nodes, such that: (i) the edges incident to nodes in $\left\{1,2,\dots,n\right\}$ only, are not included in $\mathcal{G}_{\text{INC}}$; (ii) each of the connected components of the network $\mathcal{G}$ obtained by adding the edges of  $\mathcal{G}_{\text{INC}}$ to $\mathcal{G}_{\text{EXC}}$ has a non-empty intersection with $\left\{1,2,\dots,n\right\}$. 

 To sample from $p(\mathcal{G}_{\text{EXC}}, \alpha \mid  \mathcal{G}_{\text{INC}}, D_{\text{INC}})$, we propose constructing a Gibbs sampler based on the full conditionals
\begin{displaymath}
p(\alpha\mid \mathcal{G}_{\text{EXC}}, \mathcal{G}_{\text{INC}}, D_{\text{INC}}) \quad \text{and} \quad p(\mathcal{G}_{\text{EXC}}\mid \alpha, \mathcal{G}_{\text{INC}}, D_{\text{INC}}).
\end{displaymath}
The explicit form of such conditionals will be given by the modelling choices for $p(\alpha)p(\mathcal{G} \mid \alpha)$ and the constrains imposed by $p(I \mid \mathcal{G},\eta )$. In an analogous manner, to
 sample from the posterior for $\gamma$, we impute $Y_{\text{EXC}}$ and construct a Gibbs sampler based on the full conditionals
\begin{displaymath}
p(\gamma \mid Y_{\text{EXC}},Y_{\text{INC}}, \mathcal{G}_{\text{INC}},\mathcal{G}_{{\text{EXC}}}) \quad \text{and} \quad 
p(  Y_{\text{EXC}} \mid  \gamma,Y_{\text{INC}}, \mathcal{G}_{\text{INC}},\mathcal{G}_{{\text{EXC}}}).
\end{displaymath}
Because we assume a MRF for $p(Y \mid \mathcal{G}, \gamma)$, sampling from the full conditional 
\begin{displaymath}
p(\gamma \mid Y_{\text{EXC}},Y_{\text{INC}}, \mathcal{G}_{\text{INC}},\mathcal{G}_{{\text{EXC}}})
\end{displaymath}
usually involves dealing with a distribution with an intractable partition function (See \cite{Moller}); this will be discussed with more detail in Section \ref{SubSec:BayesianComputation}.




\section{Illustration of Framework}\label{Sec:Results}

\subsection{Model Specification}\label{Se:ModSpec}
We now present the specific choices we made for these distributions: For $\mathcal{G}$ an Erd\"{o}s-R\'{e}nyi model \cite{ErdosRenyi} was assumed, with a single probability of inclusion $\alpha \in (0,1)$. A Beta$(\omega_1,\omega_2)$ was used as prior for $\alpha$. Our specification for $p(I \mid \mathcal{G},\eta)$ is an RDS with $\eta=(m,s,n)$ (\emph{i.e.}, RDS is specified with $m$ coupons per wave, $s$ seeds and sample size $n$). For the vector of responses $Y$, we assumed the Markov Random Field (MRF) implied by:
\begin{equation}
P(Y=y\mid \mathcal{G},\gamma)\propto \exp(\gamma_0 V_0+ \gamma_1 V_1),
\end{equation}
here
\begin{displaymath}
V_0=\sum_{i=1}^{N} y(i)\quad \text{and} \quad V_1=\sum_{\left\{ (i,j) \mid A_{\mathcal{G}}(i,j)=1\right\}}y(i)y(j),
\end{displaymath}
where $A_{\mathcal{G}}$ denotes the adjacency matrix for $\mathcal{G}$ and $\gamma=(\gamma_0,\gamma_1)$.
As suggested in \cite{Moller}, we assume a uniform prior on $\gamma$ of the form:
\begin{displaymath}
p(\gamma) \propto \mathbb{I}_{[\min \gamma_0, \max \gamma_0]\times[0, \max \gamma_1]}(\gamma).
\end{displaymath}

We adopt these distributions for the sake of concreteness; they are not essential to our methodology. Any sampling design that propagates through a social network could be used to specify $p(I \mid \mathcal{G},\eta)$. In principle, we could use any random graph model for $p(\mathcal{G} \mid \alpha)$, as long as:
\begin{enumerate}
\item The density of the random graph model can be computed efficiently.
\item It is feasible to marginalise efficiently, this is, to obtain the distribution of $\mathcal{G}_{\text{INC}}$ given any realization of $I$.
\end{enumerate}
For instance, one may also consider the following prior:
\begin{eqnarray}
\varphi_1, \dots, \varphi_N & \sim & B(\alpha_1,\alpha_2) \label{Eq:PP1}\\
A_{\mathcal{G}}(i,k) & \sim &\text{Ber}(\varphi_i \varphi_k),                        \label{Eq:PP2}
\end{eqnarray} 
which is inspired by the model proposed by \cite{PerryWolfe}.

\subsection{Bayesian Computation}\label{SubSec:BayesianComputation}

In Section \ref{sec-inference}, we proposed a general strategy for sampling from the posterior for $Q_\bigstar(\alpha,\gamma)$ (or $Q_\circ(\mathcal{G}_{\text{INC}}, Y_{\text{INC}})$) conditional on $(\mathcal{G}_{\text{INC}}, Y_{\text{INC}}, D_{\text{INC}})$. Such strategy is based on a Bayesian model averaging (BMA) approach (Equation \ref{Eq:BMA4RDS}) where samples from the mixing distribution and from the posterior for $\gamma$ given $(\mathcal{G}_{\text{INC}}, \mathcal{G}_{\text{EXC}},Y_{\text{INC}})$ are obtained via Gibbs-Sampler schemes. In this section, we discuss how to sample from the full conditionals implied by these schemes for the choices outlined in Section \ref{Se:ModSpec}. Since we will often deal with regimes where $N$ is at least twice bigger than $n$, and therefore, the number of entries of the adjacency matrix that we do not fully observe will be much larger than the number of entries available from $\mathcal{G}_{\text{INC}}$, we opted for proposals such that encourage imputations for $(\mathcal{G}_{\text{EXC}},Y_{\text{EXC}})$ such that they do not differ too much from $(\mathcal{G}_{\text{INC}},Y_{\text{INC}})$ from a qualitative point of view.

We first introduce some notation: Let $\mathcal{V}_\text{INC}$ and $\mathcal{V}_\text{EXC}$ the set nodes of $\mathcal{G}_{\text{INC}}$ and its complement. Let $\bar{Y}_{\text{INC}}$ denote the sample mean computed from the entries of $Y_{\text{INC}}$. Denote by $A_{\mathcal{H}}$ the submatrix of $A_{\mathcal{G}}$ (the adjacency matrix for $\mathcal{G}$) obtained by taking only the rows associated to elements of $ \mathcal{V}_\text{INC}$. We denote by $b_O$ the number of zeros in submatrix of $A_{\mathcal{G}}$ obtained by taking the rows associated to $ \mathcal{V}_\text{INC}$ and the columns associated to $ \mathcal{V}_\text{EXC}$. Denote by $j_{I}$ the number of edges included in $\mathcal{G}_{\text{INC}}$. Let $\alpha_{\text{LOW}}$ and $\alpha_{\text{UP}}$ denote, respectively, the minimum and maximum values for the density of that submatrix, given the constrains imposed by $D_{\text{INC}}$.

To sample from $p(\mathcal{G}_{\text{EXC}}\mid \alpha, \mathcal{G}_{\text{INC}}, D_{\text{INC}})$ we implemented a Metropolis step based on a mixture of four kernels. The first kernel corresponds to a proposal where a vertex $v\in \mathcal{V}_\text{INC}$ is picked uniformly at random, and all the edges connecting it to elements of $\mathcal{V}_\text{EXC}$ are re-wired, so the number of neighbours of $v$ belonging to $\mathcal{V}_\text{EXC}$ remains constant. The Metropolis ratio implied by these choices has the form
\begin{displaymath}
H^{(t)}= \frac{p\left(I\mid  \mathcal{G}_{\text{INC}},\mathcal{G}_{\text{EXC}}^{(t)} ,\eta \right)}{p\left(I\mid  \mathcal{G}_{\text{INC}},\mathcal{G}_{\text{EXC}}^{(t-1)}, \eta \right)},
\end{displaymath}
where $p(I\mid \mathcal{G}_{\text{INC}},\mathcal{G}_{\text{EXC}}^{(\cdot)}, \eta)$ is the value of Equation \ref{Eq:DistCondRDS} that results from using the imputation of the network implied by $\mathcal{G}_{\text{EXC}}^{(\cdot)}$. Note that, this move keeps the number of edges constant, therefore the terms corresponding to the Erd\"{o}s-R\'{e}nyi probability mass function cancel out. Clearly, the move that reverses the proposed one implies picking the same $v\in \mathcal{V}_\text{INC}$. By conditioning on this event, the proposal becomes a uniform over the subsets of $\mathcal{V}_\text{EXC}$ that have as many elements as v has neighbours in $\mathcal{V}_\text{EXC}$. This last statement implies that the terms corresponding to the proposal also cancel out. The second and third kernels should be seen as dual: the second kernel corresponds to the proposal where an edge connecting to vertices $(v,w)\in \mathcal{V}_\text{INC}\times \mathcal{V}_\text{INC}$ is chosen uniformly at random and then substituted by two edges, each of them connecting a different element of $\left\{v,w\right\}$ with an element of $\mathcal{V}_\text{EXC}$ (not necessarily the same one) picked uniformly at random. The third kernel allows for the opposite move: it takes two vertices $(v,w)\in \mathcal{V}_\text{INC}\times \mathcal{V}_\text{INC}$ such that, each of them has at least one edge connecting it to an element of $\mathcal{V}_\text{EXC}$, then, two of such edges are chosen (one incident to $v$ and one incident to $w$) uniformly at random and then replaced by an edge connecting $v$ and $w$. Let $h_I^{(\cdot)}$ be the number of edges of the form $(v,w)\in \mathcal{V}_\text{INC}\times \mathcal{V}_\text{INC}$ that are in the current version of the network due to imputation (\emph{i.e.}, these edges were not observed). The Metropolis ratio corresponding to the second kernel is of the form
\begin{displaymath}
H^{(t)}= \frac{\alpha}{1-\alpha} \times \frac{ p\left(I\mid  \mathcal{G}_{\text{INC}},\mathcal{G}_{\text{EXC}}^{(t)} ,\eta \right)}{ p\left(I\mid  \mathcal{G}_{\text{INC}},\mathcal{G}_{\text{EXC}}^{(t-1)}, \eta \right)} \times \frac{h_I^{(t)} \binom{b_O^{(t)}}{2}}{ \left( \frac{n(n-1)}{2}-j_I-h_I^{(t)}+1 \right) \binom{n(N-n)-b_O^{(t)}+2}{2}  },
\end{displaymath}
while the Metropolis ratio for the third kernel is given by
\begin{displaymath}
H^{(t)}= \frac{1-\alpha}{\alpha} \times \frac{ p\left(I\mid  \mathcal{G}_{\text{INC}},\mathcal{G}_{\text{EXC}}^{(t)} ,\eta \right)}{ p\left(I\mid  \mathcal{G}_{\text{INC}},\mathcal{G}_{\text{EXC}}^{(t-1)}, \eta \right)} \times \frac{ \left( \frac{n(n-1)}{2}-j_I-h_I^{(t)} \right) \binom{n(N-n)-b_O^{(t)}}{2}  }{(h_I^{(t)}+1) \binom{b_O^{(t)}+2}{2}} .
\end{displaymath}
The fourth kernel corresponds to the proposal where the submatrix of $A_{\mathcal{G}}$ with rows and columns associated to $\mathcal{V}_\text{EXC}$ is imputed using independent draws from a Bernoulli with probability of success $\alpha$. This proposal implies the Metropolis ratio
\begin{displaymath}
H^{(t)}= \frac{p\left(I\mid  \mathcal{G}_{\text{INC}},\mathcal{G}_{\text{EXC}}^{(t)} ,\eta \right)}{p\left(I\mid  \mathcal{G}_{\text{INC}},\mathcal{G}_{\text{EXC}}^{(t-1)}, \eta \right)},
\end{displaymath}
since the terms corresponding to the proposal and those corresponding to the random graph distribution (given $\alpha$) cancel out.

To sample from $p(\alpha\mid \mathcal{G}_{\text{EXC}}, \mathcal{G}_{\text{INC}}, D_{\text{INC}})$ we implemented a Metropolis step based on a mixture of two kernels. The first kernel corresponds to a Beta$(\kappa_1,\kappa_2)$ proposal, where  
\begin{displaymath}
\kappa_1=0.5+\sum_{i=1}^n \sum_{j=i+1}^N A_{\mathcal{H}}(i,j) \quad \text{and}  \quad \kappa_2=0.5+\left[ \frac{(n-1)n}{2}+ n(N-n)\right] -\sum_{i=1}^n \sum_{j=i+1}^N A_{\mathcal{H}}(i,j).
\end{displaymath}
The intuition behind this proposal is the following: this distribution can be understood as the posterior for $\alpha$ implied by the Jeffreys prior, Beta$(0.5,0.5)$, and the counts for presence and absence of edges in the submatrix of $A_{\mathcal{G}}$ obtained by taking only the rows corresponding to elements of $ \mathcal{V}_\text{INC}$. By adopting this proposal, the Metropolis ratio required for computing the acceptance probability for a new move $\alpha^{(t)}$ takes the form:
\begin{equation}\label{Eq:PropAlphaKer}
H^{(t)}= \frac{(\alpha^{(t)})^{h_{\mathcal{G}}-\frac{1}{2}}(1-\alpha^{(t)})^{\frac{N(N-1)}{2}-h_{\mathcal{G}}-\frac{1}{2}}}{(\alpha^{(t-1)})^{h_{\mathcal{G}}-\frac{1}{2} }(1-\alpha^{(t-1)})^{\frac{N(N-1)}{2}-h_{\mathcal{G}} -\frac{1}{2} }}\times\frac{(\alpha^{(t-1)})^{\kappa_1-1}(1-\alpha^{(t-1)})^{\kappa_2-1}}{(\alpha^{(t)})^{\kappa_1-1}(1-\alpha^{(t)})^{\kappa_2-1}},
\end{equation}
where $h_{\mathcal{G}}$ denotes the number of non-zero entries of the upper-triangular part of $A_{\mathcal{G}}$. The second kernel corresponds to a Beta$(\zeta_1,\zeta_2)$  proposal, where  $\zeta_1$ and $\zeta_2$ are such that, the $5$ and $95$\% quantiles of this Beta distribution are equal to $\alpha_{\text{LOW}}$ and $\alpha_{\text{UP}}$, respectively. The Metropolis ratio implied by this proposal has the same form as the one shown in Equation \ref{Eq:PropAlphaKer}, one just needs to replace $(\kappa_1,\kappa_2)$ by $(\zeta_1,\zeta_2)$.

To sample from $p(  Y_{\text{EXC}} \mid  \gamma,Y_{\text{INC}}, \mathcal{G}_{\text{INC}},\mathcal{G}_{{\text{EXC}}})$, we use a Metropolis step with proposal $q(Y_{\text{EXC}})$ given by a multivariate Bernoulli with all marginal probabilities of success equal to $\bar{Y}_{\text{INC}}$ and such that its components are independent. To accept a new proposed value $Y_{\text{EXC}}^{(t)}$, we compute the following Metropolis ratio:
\begin{displaymath}
H^{(t)}= \frac{\exp(\gamma_0 V_0^{(t)}+\gamma_1 V_1^{(t)})}{\exp(\gamma_0 V_0^{(t-1)} + \gamma_1 V_1^{(t-1)})}\times\frac{(\bar{Y}_{\text{INC}})^{c^{(t-1)}}(1-\bar{Y}_{\text{INC}})^{(N-n)-c^{(t-1)}}}{(\bar{Y}_{\text{INC}})^{c^{(t)}}(1-\bar{Y}_{\text{INC}})^{(N-n)-c^{(t)}}},
\end{displaymath}
where $V_0^{(\cdot)}$ and $V_0^{(\cdot)}$ are, respectively, the values of $V_0$ and $V_1$ implied by $Y_{\text{EXC}}^{(\cdot)}$. Here $c^{(\cdot)}$ denotes the number of non-zero entries of $Y_{\text{EXC}}^{(\cdot)}$.

To sample from $p(\gamma \mid Y_{\text{EXC}},Y_{\text{INC}}, \mathcal{G}_{\text{INC}},\mathcal{G}_{{\text{EXC}}})$ using a Metropolis step, we have to deal with the presence of intractable partition functions on a Metropolis ratio. To tackle this task, we follow the approach proposed by \cite{AndrieuRob}, which is based on using an unbiased estimate of the density function within the Metropolis ratio; more specifically, we implemented the Annealed Importance Sampling (AIS) algorithm (See \cite{Neal} and Section 3.2 of \cite{Salak}) to compute unbiased estimates of the partition function whenever needed. We describe the AIS algorithm briefly in Appendix \ref{App:AIS}. We used a mixture of kernels to update $\gamma$. Each element of the mixture corresponds to a random walk (as described in Section 7.5 of \cite{RoberCase}) for one of the components of $\gamma$; more specifically, the proposals are based on perturbations of $\gamma_0^{(t)}$ (or $\gamma_1^{(t)}$) generated from a uniform on $[-\upsilon,\upsilon]$. We consider random walks with reflection on the boundary. These choices imply Metropolis ratios of the form
\begin{displaymath}
H^{(t)}= \frac{\exp(\gamma_0^{(t)} V_0+\gamma_1 V_1)}{\exp(\gamma_0^{(t-1)} V_0+\gamma_1 V_1)}\times\frac{\hat{Z}(\gamma_0^{(t-1)},\gamma_1)   }{\hat{Z}(\gamma_0^{(t)},\gamma_1)},
\end{displaymath}
where $\hat{Z}(\gamma_0^{(\cdot)},\gamma_1)$ denotes the estimate of the partition function computed by AIS and implied by $\gamma_0^{(\cdot)}$. Note that we are showing the Metropolis ratio for an update for $\gamma_0$, the ratio for performing an update on $\gamma_1$ is obtained \emph{mutatis mutandis}.

\subsection{Simulation Study} \label{Sec:SimulStudy}
We conducted a simulation study to gain better understanding of the performance of our method. First we considered regimes where the random graph model $p(\mathcal{G}\mid \alpha)p(\alpha)$ matches the mechanism that generated the data. To evaluate performance of point estimators, Monte Carlo estimates of the bias and the mean square error (MSE) were computed;  to evaluate confidence intervals and credible regions, the frequency of coverage was used. The regimes for the simulation were determined by the following factors:
\begin{enumerate}
\item The density of the underlying network. We considered the values $0.005$ and $0.01$.
\item The size of the underlying network. We considered the values $500$, $1000$ and $2000$.
\item The dependence among the entries of the response vector induced by the Markov random field. We considered $\gamma_1\in\left\{ 0.001,0.25 \right\}$. The value $\gamma=0.001$ defines regimes with low dependence among the responses; in contrast, $\gamma=0.25$ defines the regimes with high dependence. For each regime, we specify $\gamma_0$ by finding the value $\gamma_0^*$ such that Equation \ref{Eq:TheQDef} holds, given $(\alpha,\gamma_1,Q_\bigstar)$.
\end{enumerate}
The sample size was set as $150$. An Erd\"{o}s-R\'{e}nyi model was used to generate the random graph data for all regimes. We set $Q_\bigstar=0.2$ for all scenarios. Our method was implemented using the distributions described in Section \ref{Se:ModSpec}. We compared our methodology to the Volz-Heckathorn estimator and the corresponding Bootstrap confidence interval. Results are summarised in Table \ref{Tab:SimRightSpec}. We observed that, for most regimes, the estimator for $Q_\bigstar$ implied by our method had less bias than the VH estimator. In terms of MSE, the VH estimator outperforms the Bayesian estimator when the dependence among the responses is low; in contrast, the estimator implied by our method outperforms the VH estimator when the dependence among the responses is high. 

\begin{table}
\caption{\label{Tab:SimRightSpec}Average bias, $Q - \hat{Q}$, and MSE for the Bayesian and non-model based approach. For the Bayesian method, the point estimator (either $\hat{Q}_\bigstar$ or $\hat{Q}_\circ$) is given by the posterior mean. We compared this summary to the Volz-Heckatorn (VH) estimator. The simulation scenarios are given by: Density of the underlying network, dependence among the components of the response vector and the size of the underlying network. The sample size was set as $150$ for all scenarios. $100$ simulations were performed for each scenario. In all cases an Erd\"{o}s-R\'{e}nyi model is used to generate the data. For each simulation, the BMA was implemented using $4$ samples from the mixing distribution; for each of these samples, an MCMC was run using $1,000$ for burn-in and $50$ posterior samples.}
\centering
\fbox{%
\begin{tabular}{| c c c c  c | c  c c c |l |}
\hline
Regime & $\alpha$ & $\gamma_0$ & $\gamma_1$ & $N$ & Bias $\hat{Q}_\bigstar$ & MSE $\hat{Q}_\bigstar$ & Bias $\hat{Q}_\circ$ & MSE $\hat{Q}_\circ$ & Method \\
\hline
 1  &  0.005    &  -0.6816 & 0.001  &   500    &  -0.0057  &   0.0022  &  0.0093 & 0.0001 & Bayes  \\
 1  &  0.005    &  -0.6816 & 0.001  &   500    &  -0.0231   &  0.0022   &  -0.0155 & 0.0006 & VH  \\
 2  &  0.005    &  -0.6921 & 0.001  &  1000   &  -0.0026   &  0.0032  &  0.0160 & 0.0002 & Bayes  \\
 2  &  0.005    &  -0.6921 & 0.001  &   1000   &  -0.0238   &  0.0025  & -0.0025 & 0.0001 & VH  \\
 3  & 0.005    &   -0.6927 & 0.001  &    2000   &  -0.0107   &  0.0044   &  0.0154 & 0.0007 & Bayes  \\
 3  &  0.005    &  -0.6927 & 0.001  &   2000   &  -0.0082   &   0.0021  &  -0.0042 & 0.0001 & VH  \\
 4  &  0.005    &  -0.8271 & 0.250  &   500     &  -0.0212   &   0.0025   &  0.0007 & 0.0005 & Bayes \\
 4  &  0.005    &  -0.8271 & 0.250  &   500     &  -0.0459   &   0.0028   &  -0.0026 & 0.0005 & VH  \\
 5  &  0.005    &  -1.0165 & 0.250  &   1000   &  0.0128   &   0.0017   &  0.0023 & 0.0007 & Bayes  \\
 5  &  0.005    &  -1.0165 & 0.250  &   1000   &  -0.0202   &   0.0026   &  0.0034 & 0.0008 & VH  \\
 6  &  0.005    &  -1.3409 & 0.250  &   2000   &  0.0167  &   0.0016   &  0.0070 & 0.0014 & Bayes  \\
 6  &  0.005    &  -1.3409 & 0.250  &   2000   &  -0.0201   &   0.0018   &  0.0324 & 0.0012 & VH  \\
 7  &  0.01      &  -0.6858 & 0.001  &   500     &  0.0022   &   0.0027   &  0.0212 & 0.0008 & Bayes  \\
 7    &  0.01    &  -0.6858 & 0.001  &   500    &   0.0048   &   0.0009   &  0.0070 & 0.0001 & VH  \\
 8    &  0.01    &  -0.6922 & 0.001  &   1000     &  0.0022   &   0.0044   &  0.0193 & 0.0011 & Bayes \\
 8    &  0.01    &  -0.6922 & 0.001  &   1000     &  0.0051   &   0.0011   &  0.0078 & 0.0004 & VH  \\
 9  &  0.01      &  -0.6993 & 0.001   &   2000     &  -0.0020   &   0.0027   &  0.0143 & 0.0004 & Bayes  \\
 9  &  0.01      &  -0.6993 & 0.001   &   2000    &  -0.0472   &   0.0022   &  0.0132 & 0.0006 & VH  \\
 10  &  0.01    &  -1.0049 & 0.250  &    500    &  -0.0395   &   0.0016   &  0.0012 & 0.0003 & Bayes  \\
 10  &  0.01    &  -1.0049 & 0.250  &    500    &  -0.0478  &   0.0023   &  -0.0143 & 0.0004 & VH  \\
 11  &  0.01    &  -1.3497 & 0.250  &    1000   &  -0.0338   &   0.0026   & 0.0078  & 0.0004 & Bayes  \\
 11  &  0.01    &  -1.3497 & 0.250  &    1000    &  -0.0487  &   0.0033  &  0.0073 & 0.0012 & VH  \\
 12  &  0.01    &  -2.1525 & 0.250  &    2000     &  -0.0208   &   0.0016   & -0.0031 & 0.0007 & Bayes \\
 12  &  0.01    &  -2.1525 & 0.250  &    2000     &  -0.0402  &   0.0018   &  -0.0013 & 0.0026 & VH  \\
\hline
\end{tabular}}
\end{table}

\begin{figure}
\centering
\includegraphics[width=3.5in]{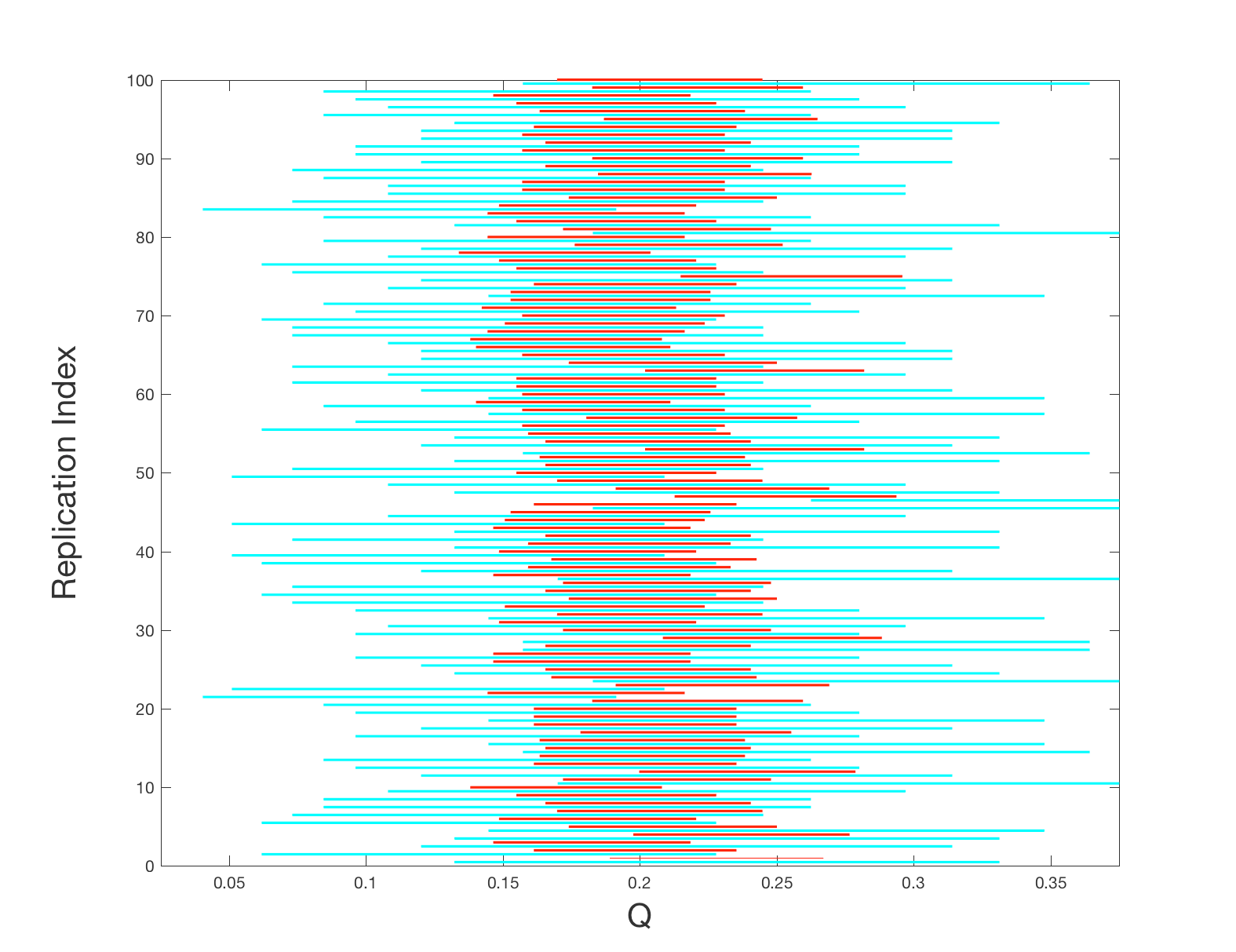}
\caption{95\% credible intervals (red) \emph{vs.} 95\% bootstrap confidence intervals (cyan) for $Q_\bigstar$. For all data sets, the credible interval is plotted on top of the corresponding confidence interval. 100 Monte Carlo data sets were used to obtain this plot. We observed that our method produces intervals that are at least half shorter in average while having slightly higher coverage.}\label{Fig:CIntSingle}
\end{figure}

\begin{figure}[t!]
\centering
 \includegraphics[width=4in]{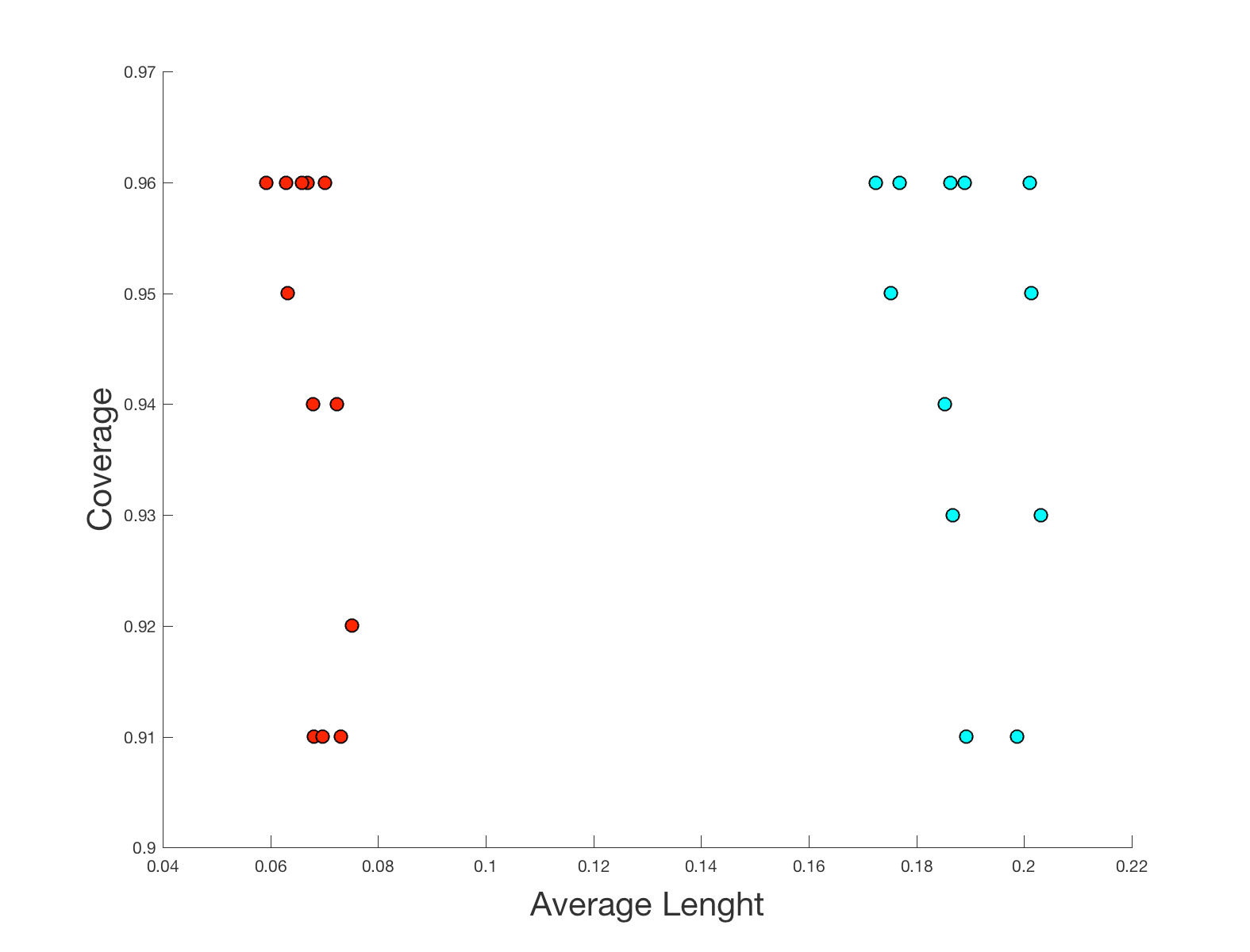}
\caption{Average length \emph{vs.} coverage of credible sets (red) and confidence intervals (cyan) for $Q_\bigstar$. These results correspond to the regimes listed in Table \ref{Tab:SimRightSpec}. }\label{F:CIntMultiple}
\end{figure}

For a given regime (namely, $N=1000$, $\alpha=0.001$, $\gamma=(-0.6927,0.001)$), we plotted the $95$ per cent credible intervals associated to each simulation and the corresponding $95$ per cent confidence intervals implied by bootstrapping Volz-Heckathorn (Figure \ref{Fig:CIntSingle}). While both procedures generate estimates with similar bias and coverage, our method produces intervals that are at least half shorter in average. We also plotted the coverage against the average length of the (confidence or credible) interval for each method considering a the same set of regimes used in Tables \ref{Tab:SimRightSpec}. Results are displayed in Figure \ref{F:CIntMultiple}.

\begin{figure}[t!]
\begin{center}
 \includegraphics[width=5.5 in]{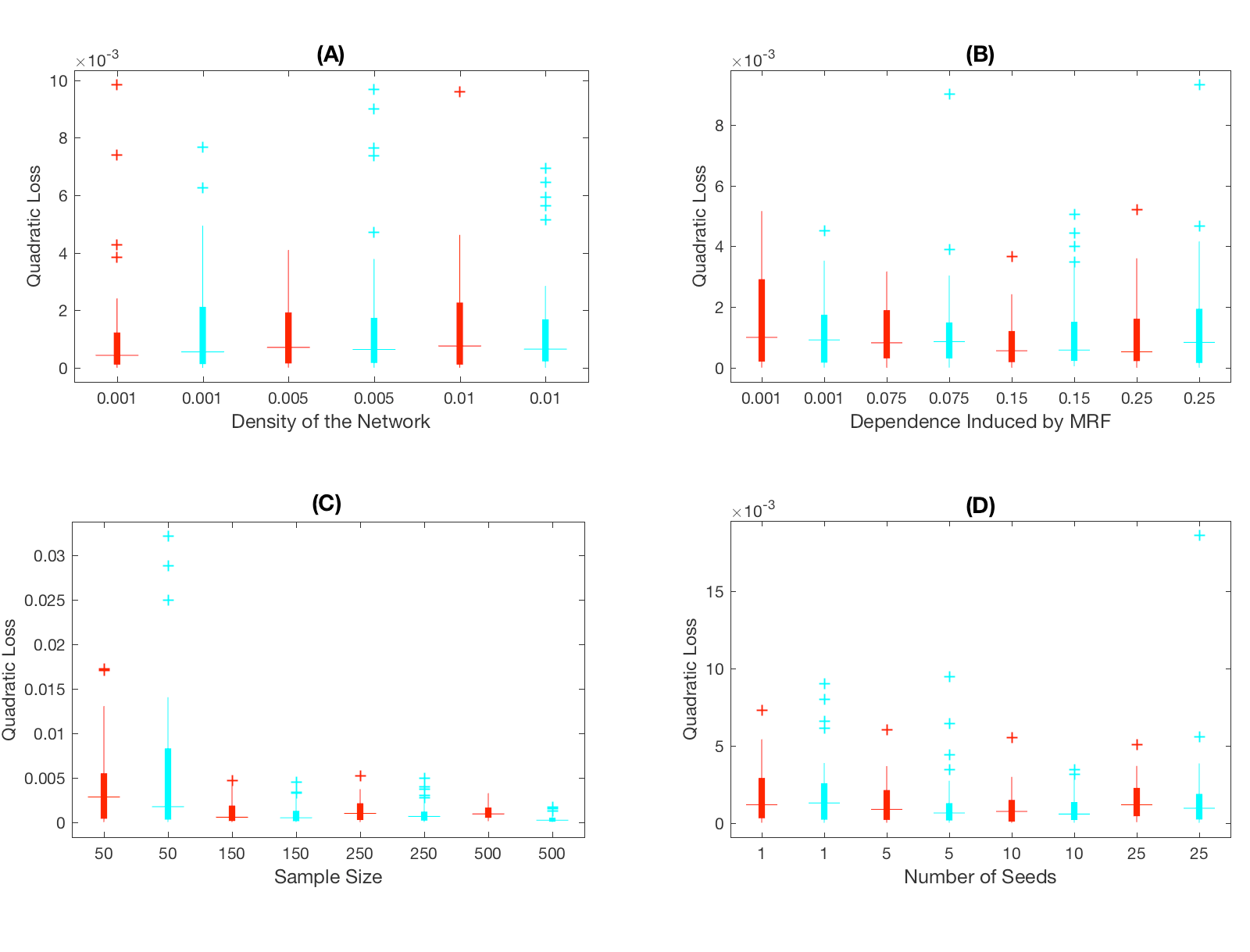}
\end{center}
\caption{ In these plots we explore the behaviour of  our method (red) and the VH estimator (cyan) in terms of the distribution of the quadratic loss. For all the following scenarios, we specified $Q_\bigstar=0.2$, $N=1000$  and $m=3$. In panel (A), we consider $\alpha \in \left\{ 0.001, 0.005, 0.01 \right\}$ and set $\gamma_1=0.15$, $n=150$ and $s=1$. In panel (B), we consider $\gamma_1 \in \left\{ 0.001, 0.075, 0.15, 0.25   \right\}$ and set $\alpha=0.005$, $n=150$ and $s=1$. In panel (C), we consider $n \in \left\{ 50,150, 250, 500 \right\}$ and set $\alpha=0.005$, $\gamma_1=0.15$ and $s=1$. In panel (D), we set $s \in \left\{ 1, 5, 10, 25 \right\}$ and set $\alpha=0.005$, $\gamma_1=0.15$ and $n=150$.}\label{Fig:MultRegimesPlot}
\end{figure}

\begin{figure}[t!]
\begin{center}
 \includegraphics[width=5.5 in]{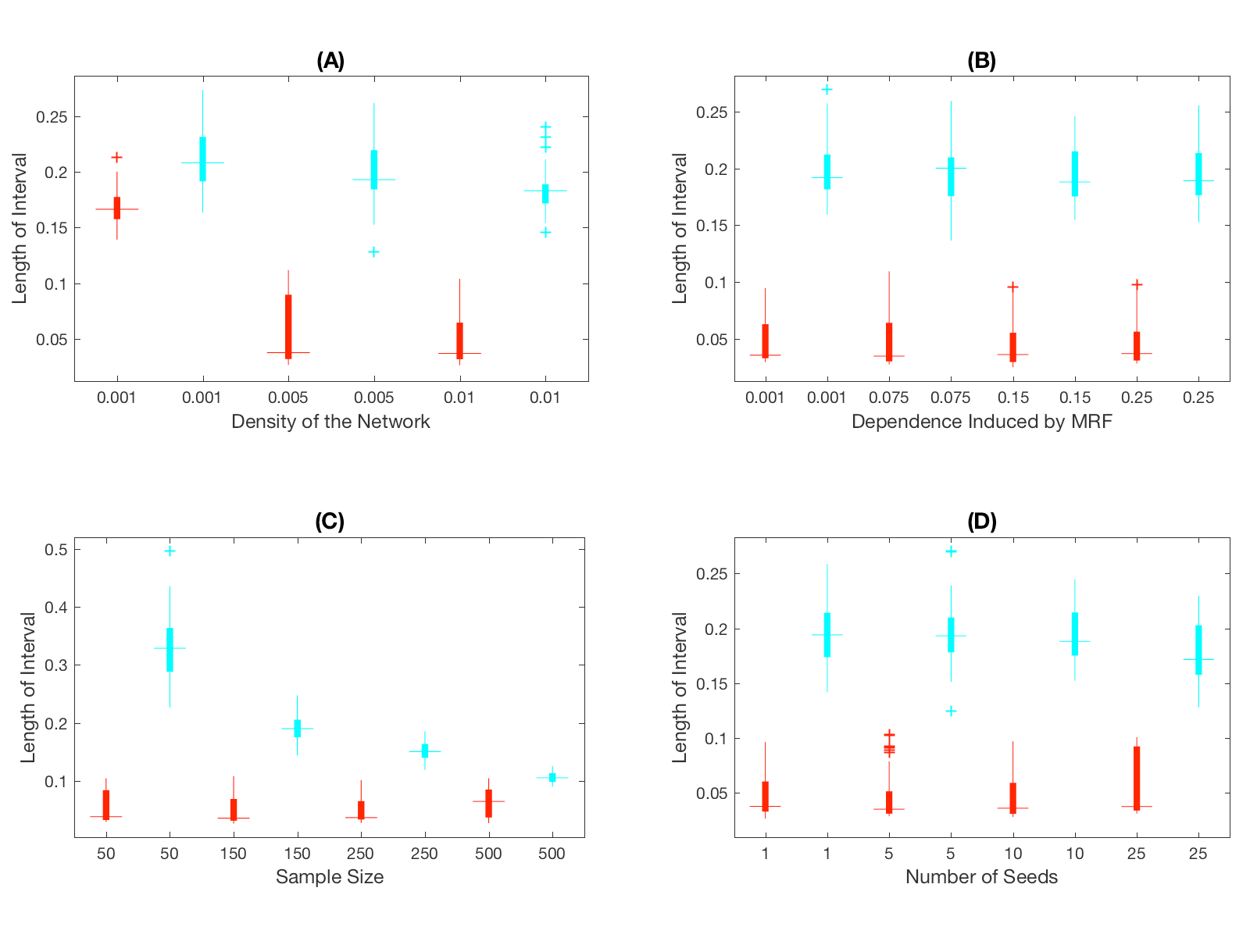}
\end{center}
\caption{ In these plots we explore the behaviour of  our method (red) and the VH estimator (cyan) in terms of the distribution of the length of the $95$\% credible region (Bayesian method) or the $95$\% confidence interval (VH). For all the following scenarios, we specified $Q_\bigstar=0.2$, $N=1000$  and $m=3$. In panel (A), we consider $\alpha \in \left\{ 0.001, 0.005, 0.01 \right\}$ and set $\gamma_1=0.15$, $n=150$ and $s=1$. In panel (B), we consider $\gamma_1 \in \left\{ 0.001, 0.075, 0.15, 0.25   \right\}$ and set $\alpha=0.005$, $n=150$ and $s=1$. In panel (C), we consider $n \in \left\{ 50,150, 250, 500 \right\}$ and set $\alpha=0.005$, $\gamma_1=0.15$ and $s=1$. In panel (D), we set $s \in \left\{ 1, 5, 10, 25 \right\}$ and set $\alpha=0.005$, $\gamma_1=0.15$ and $n=150$. For all these regimes, the Monte Carlo estimates of the coverage were close to the nominal value, as in Figure \ref{F:CIntMultiple}. }\label{Fig:MultRegimesPlot_Intervals}
\end{figure}

We performed a second set of simulations, again specifying  $Q_\bigstar=0.2$. We compared our method to the VH estimator in terms of the distribution of quadratic loss $(Q_\bigstar-\hat{Q}_\bigstar)^2$ and the distribution of the length of credible\slash confidence intervals. For this simulation we set $N=1000$ and $m=3$ for all regimes; $\gamma_1=0.15$ unless stated otherwise. The regimes we considered for these simulations are defined in terms of density of the underlying network ($\alpha \in \left\{ 0.001, 0.005, 0.01 \right\}$, panel (A)), dependence among the responses ($\gamma_1 \in \left\{ 0.001, 0.075, 0.15, 0.25   \right\}$, panel (B)), sample size ($n \in \left\{ 50,150, 250, 500 \right\}$, panel (C)) and the number of seeds ($s \in \left\{ 1, 5, 10, 25 \right\}$, panel (D)). Results from this simulation are summarised in Figures \ref{Fig:MultRegimesPlot} and \ref{Fig:MultRegimesPlot_Intervals}.  As in the previous simulation experiment, we observed that, in terms of quadratic loss, there are no substantial differences between our method and VH. We also observed that, the credible intervals were much shorter than the corresponding confidence intervals, while keeping the coverage close to the nominal value.

\subsection{When the Prior on $\mathcal{G}$ is Misspecified} 
\begin{figure}[t!]
\begin{center}
 \includegraphics[width=4.5 in]{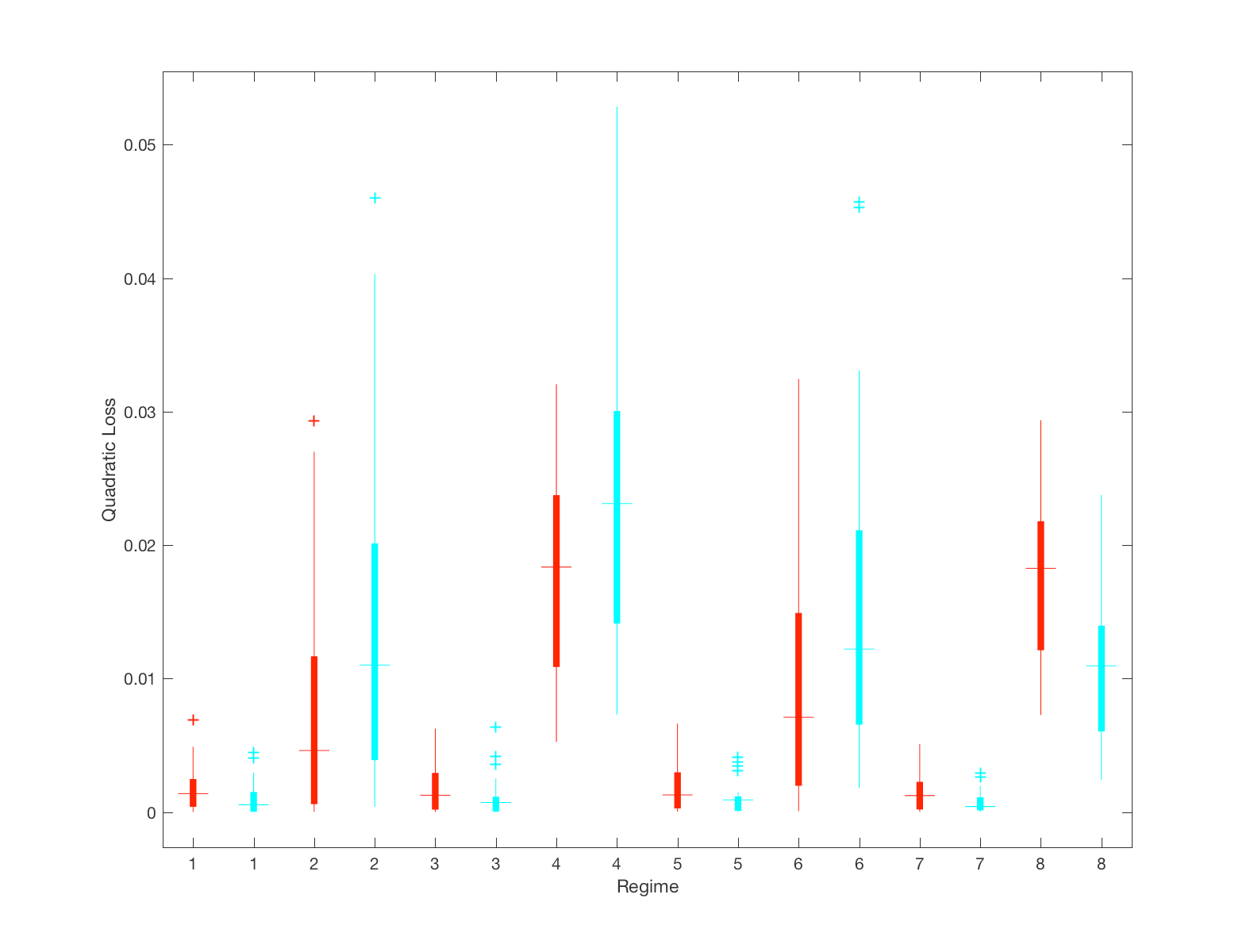}
\end{center}
\caption{ In these plots we explore the behaviour of  our method (red) and the VH estimator (cyan) in terms of the distribution of the quadratic loss for the estimation of  $Q_\bigstar$. The underlying graph is obtained from a Small-World model, while the inference procedures assumes an Erd\"{o}s-R\'{e}nyi model. The regimes for the simulation are displayed in Table \ref{Tab:SimMissReg}. The values for $\gamma_0$ were set up so $Q_\bigstar=0.2$ for all regimes. For this simulation, the size of the network was set as $N=1000$, the sample size as $n=150$, the number of referrals  as $m=3$ and the number of seeds as $s=1$.}\label{Fig:QLossSW}
\end{figure}

\begin{figure}[h!]
\centering
 \includegraphics[width=4.5in]{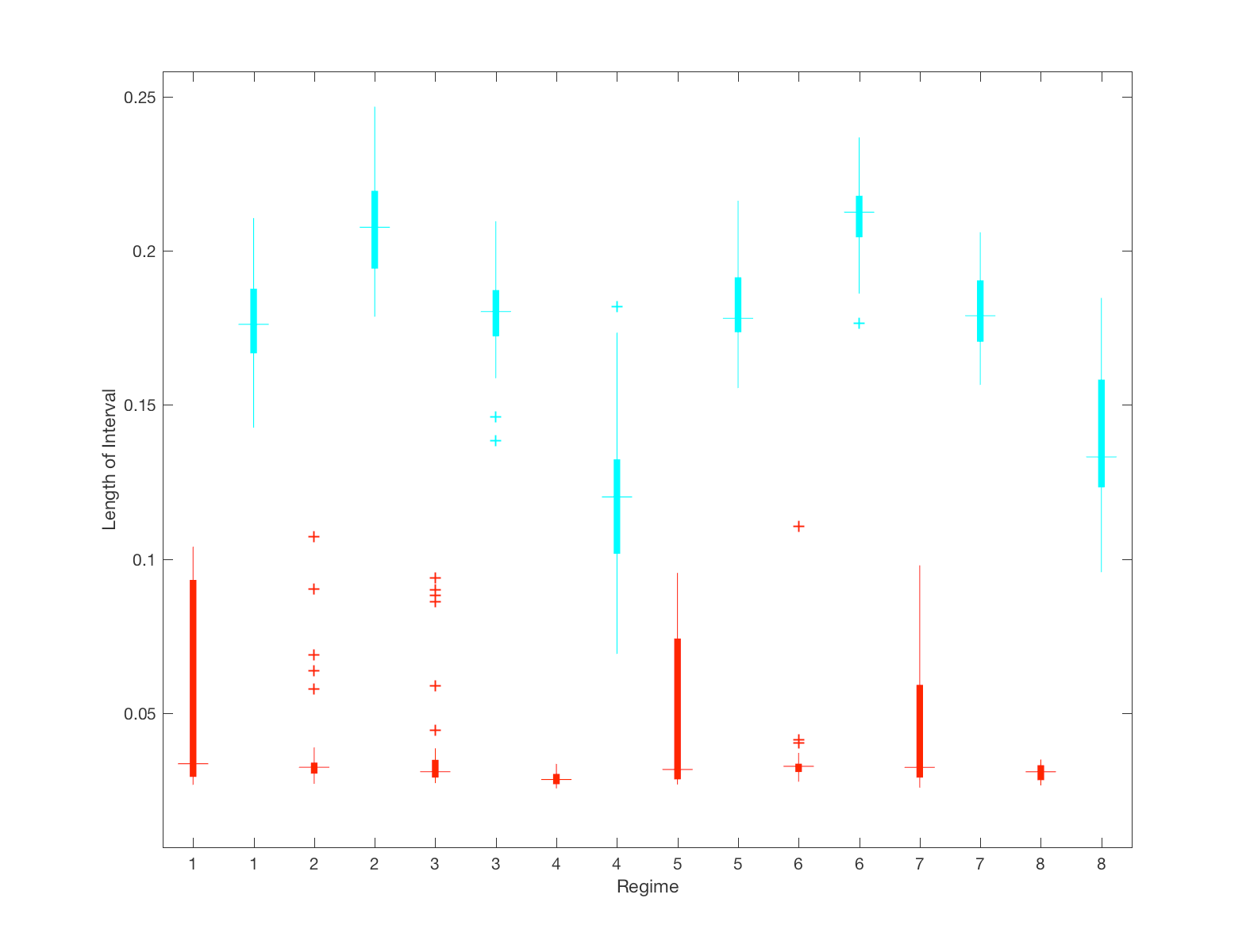}
\caption{95\% credible intervals (red) \emph{vs.} 95\% bootstrap confidence intervals (cyan) for $Q_\bigstar$. The underlying graph is obtained from a Small-World model, while the inference procedures assumes an Erd\"{o}s-R\'{e}nyi model. The regimes for the simulation are displayed in Table \ref{Tab:SimMissReg}. The values for $\gamma_0$ were set up so $Q_\bigstar=0.2$ for all regimes. For this simulation, the size of the network was set as $N=1000$, the sample size as $n=150$, the number of referrals  as $m=3$ and the number of seeds as $s=1$. }\label{Fig:CIntSW}
\end{figure}

We performed one more simulation study where we considered regimes for which the prior $p(\mathcal{G}\mid \alpha)p(\alpha)$ had different functional form from the mechanism that generated the data. We evaluated the procedures using the same criteria as in the experiments described in Section \ref{Sec:SimulStudy}. The regimes for the simulation are displayed in Table \ref{Tab:SimMissReg}; these were determined by the following factors:
\begin{enumerate}
\item The random graph model used to generate the underlying network was Small World with density $\xi\in \left\{ 0.005,0.01,0.05 \right\}$ and probability of re-wiring $\vartheta \in \left\{  0.01, 0.25\right\}$.
\item The random graph models used to fit the model were  Erd\"{o}s-R\'{e}nyi.
\item The dependence among the responses we determined by  $\gamma_1\in \left\{0.001,0.25\right\}$.
\end{enumerate}
The values for $\gamma_0$ were set up so $Q_\bigstar=0.2$ for all regimes. For this simulation, the size of the network was set as $N=1000$, the sample size as $n=150$, the number of referrals  as $m=3$ and the number of seeds as $s=1$. Again, our method was implemented using the distributions described in Section \ref{Se:ModSpec}. Results are summarised in Figures \ref{Fig:QLossSW} and \ref{Fig:CIntSW}. As in the simulations performed in Section \ref{Sec:SimulStudy}, our method gets similar results to VH in terms of quadratic loss. The regimes that prove harder for both methods, in terms of quadratic loss, are the ones with high dependence and high density of the network (Regimes 4 and 8 in Table \ref{Tab:SimMissReg}).  We also observed that the credible intervals implied by our method tend to be much shorter than the corresponding Bootstrap confidence intervals, as in the results obtained in Section \ref{Sec:SimulStudy}. For both methods and all regimes, the coverage of the intervals (credible or frequentist) were close to the nominal value. 

\begin{table}
\caption{Simulation regimes for the results in Figures \ref{Fig:QLossSW} and \ref{Fig:CIntSW}. The regimes are given by the probability of re-wiring $\vartheta$, the density of the network $\xi$, and the parameters of the Marvok random field $(\gamma_0,\gamma_1)$ \label{Tab:SimMissReg}.}
\centering
\fbox{%
\begin{tabular}{| c | c c c c |}
\hline
Regime & $\vartheta$ & $\xi$ & $\gamma_0$ & $\gamma_1$  \\
\hline
 1    &  0.01  &   0.005    &  -0.6927    &    0.001      \\
 2    &  0.01   &  0.005    &  -0.8271    &    0.25      \\
 3    &  0.01   &  0.01      &  -0.6993    &    0.001              \\
 4    &  0.01   &  0.01      &  -1.0049    &    0.25               \\
 7    &  0.25   &  0.005    &  -0.6927    &    0.001       \\
 8    &  0.25   &  0.005    &  -0.8271    &    0.25       \\
 9    &  0.25   &  0.01      &  -0.6993    &    0.001      \\
 10  &  0.25   &  0.01      &  -1.0049    &    0.25      \\
\hline
\end{tabular}}
\end{table}

 \section{Real Data}\label{Sec:RealData}
 We applied our methodology to the data derived from the study discussed in \citep{deMelPinho}. This was a large RDS study implemented in a single location, namely the community of Campinas in the state of Sao Paulo, Brazil. Since RDS was used, non-ignorability is an issue for likelihood-based inferences. The aim of the study was to infer the prevalence of HIV  among gay men in Campinas, Brazil.
   
The study comprised  $658$ men who have sex with men. The inclusion criteria used for this study, were:
            \begin{enumerate}
            \item born male;
            \item had anal or oral sex with another man or transvestite in the past six months;
            \item 14 years of age or older;
            \item reside in the Metropolitan area of Campinas.
            \end{enumerate}
RDS was implemented using $16$ seeds and a maximum of $3$ referrals per subject  (\emph{i.e.}, $m=3$). Point estimates (sample proportion and Volz-Heckathorn) and Bootstrap confidence intervals are shown in Table \ref{Tab:FreqRes}. The results shown in this table are not model-based.

We applied our method to this data set; the results are summarised in Table \ref{Tab:BayesRes}. We used the model specification presented in Section \ref{Se:ModSpec} and adopted the super-population quantity (Equation \ref{Eq:TheQDef}) as the estimand. We used a Beta$(0.5,0.5)$ as prior for $\alpha$ and an uniform on $[-3.5,1]\times[0,1]$ for $(\gamma_0,\gamma_1)$. Remember that $N$ is a tuning parameter for our model, therefore, a sensitivity analysis with respect to it is required. The model was fit with $N$ specified as two, three and four times the sample size $n$ (\emph{i.e.}, 1316,1974 and 2632). For the mixing distribution, we obtained $150$ samples after $2,500$ iterations for burn-in. We took five of those samples, uniformly at random, and computed the posterior for $(\gamma_0,\gamma_1)$ conditional on each of these samples. For this computation, we obtained $100$ posterior samples after a burn-in of $2,000$ iterations. For each pair $(\alpha,\gamma)$, a Monte Carlo version of $Q$ was computed from $100$ simulated values (Equation \ref{Eq:TheQDef}). The point estimates and credible regions we obtained for $Q$ were very stable with respect to $N$. The point estimates for $Q$ were slightly higher than the results from the naive estimator and Volz-Heckathorn. The credible regions obtained from our method were consistently shorter than the Bootstrap confidence interval associated to Volz-Heckathorn.

\begin{table}
\caption{\label{Tab:FreqRes}Point estimators (sample proportion, Volz-Heckathorn) and the corresponding $95$ per cent Bootstrap confidence intervals.}
\centering
\fbox{%
\begin{tabular}{|l | c | c |}
\hline
         & Naive & Volz-Heckathorn\\
\hline
 $\hat{Q}$ &  $0.0789$ \quad $(0.0577,0.1001)$      &  $0.0711$ \quad $(0.0466,0.0955)$  \\
\hline
\end{tabular}}
\end{table}

\begin{table}
\caption{\label{Tab:BayesRes} Summaries of the marginal posteriors corresponding to $Q$, $\alpha$ and $(\gamma_0,\gamma_1)$. These summaries are: the posterior mean and the 2.5, 50 and 97.5\% posterior quantiles. } 
\centering
\begin{tabular}{| c | c | c  c c | c |}
\hline
Parameter & mean  & $0.025$ & $0.5$ & $0.975$ &N\\
\hline
$Q$                 &  0.0795 &   0.0578   &  0.0793   &   0.0965  & 1316 \\
$\alpha$          &  0.0092 &   0.0090   &  0.0093   &   0.0100  & 1316     \\
$\gamma_0$   &  -2.3896 &   -2.9182   &  -2.5315   &  -1.1555   & 1316   \\
$\gamma_1$   &  0.1197 &   0.0011   &  0.0852 &    0.5048  & 1316  \\
\hline
$Q$   &  0.0791 &   0.0628    &  0.0784   &    0.0960   & 1974 \\
$\alpha$        &  0.0059 &   0.0056    &  0.0059   &    0.0063 & 1974  \\
$\gamma_0$         &  -2.5202 &   -2.9460    &  -2.5608   &    -1.9281  & 1974 \\
$\gamma_1$        &  0.1780 &   0.0137   &  0.0858 &    0.6476  & 1974 \\
\hline
$Q$                 &  0.0737 &   0.0592   &  0.0730   &   0.0894  & 2632 \\
$\alpha$          &  0.0043 &   0.0042   &  0.0043   &   0.0046  & 2632     \\
$\gamma_0$   &  -2.5144 &   -3.4876   &  -2.4780  &  -1.4954   & 2632   \\
$\gamma_1$   &  0.1891 &   0.0073   &  0.0908 &    0.6354  & 2632 \\
\hline
\hline
\end{tabular}
\end{table}

\subsection{Assessing Goodness of Fit}
\begin{figure}[t!]
\begin{center}
 \includegraphics[width=4 in]{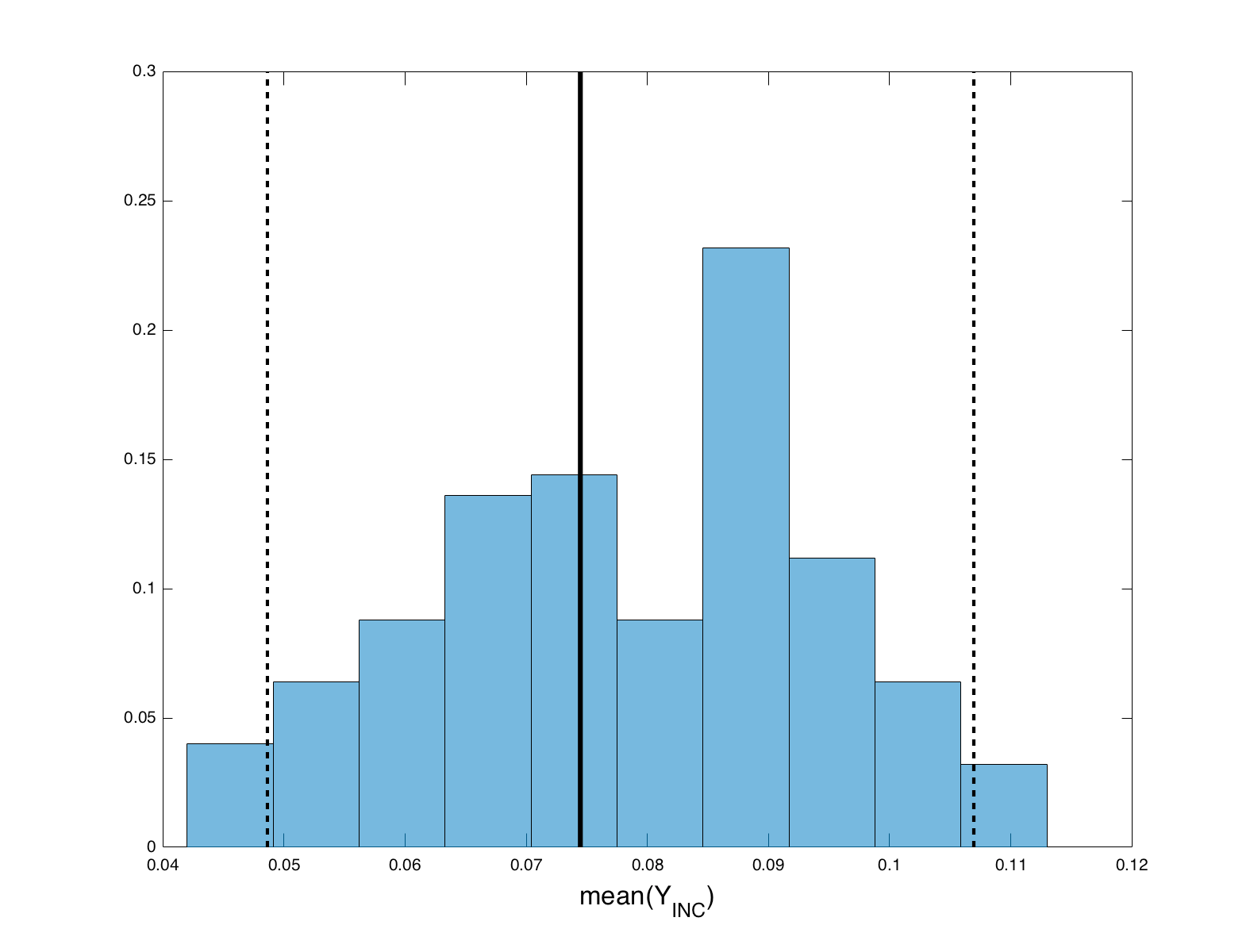}
\end{center}
\caption{Sample mean of $Y_{INC}$ (red line) compared to Monte Carlo distribution of repetitions of $Y_{INC}$ obtained via the posterior predictive distribution. This was done for a networks with $N=200$ and $\alpha=0.01$ (left) and $0.05$ (right). The same exercise was performed for $\alpha=0.1$ (left) and $0.2$ with similar results (not shown).}\label{Fig:PostPredChck}
\end{figure}

To assess the goodness of fit of out model we made use of posterior predictive checks (See \citep{GelMeng} and the appendix), \emph{i.e.}, we selected a summary of the observed data and plotted the observed value of this summary against the distribution of replicates of such summary under the posterior predictive distribution. For this case, we used the sample mean of the observed $Y$'s as the summary. Results are displayed in Figure \ref{Fig:PostPredChck}. The plot does not show evidence against the goodness of fit of our model.




\section{Discussion}\label{Sec:Discuss}

In this paper, we study statistical properties of popular network sampling mechanisms, and we developed methodology for performing inference on non-ignorable designs on a network. 
We make the following contributions: 
 We extend Rubin's notion of ignorability of the sampling mechanisms to explicitly account for social network structure; 
 We calculate the probability model for Respondent Driven Sampling (or RDS), a popular network sampling mechanism, and prove that it is not ignorable; 
 We show (empirically) that for achieving (Bayesian) interval estimates with nominal coverage, when that data has been collected using a sampling mechanism that is not-ignorable, one needs to take into account the probability model of the sampling mechanism at hand, explicitly. 

We note an important distinction between the case of non-ignorable sampling and non-ignorable missing data. On the one hand, there is largely no control on missing data mechanisms, in practice. Assumptions of ignorable missing data are not testable. The only sensible robustness check is to explore how sensitive the results of the analysis are to such assumption. On the other hand, the experimenter has typically full control on the sampling mechanism used to collect data. So the ignorability of a sampling mechanism is a condition that can (and should) be checked. In the HIV example we analyze, for instance, RDS is used. It is reasonable to assume the experimenter can write down a reasonable approximation of the probability model. And with that is hand, one can check the technical ignorability condition. In practice, the missing data and sampling mechanisms are confounded, as one only observed responses that are both included in the sample and non-missing. Nonetheless, this is an important distinction for inferential purposes. In this paper, we have assumed no missing data to elucidate the effects of non-ignorable sampling mechanisms, in theory and in practice.

The authors in  \citep{HandcGile} discuss the idea of \emph{amenable} designs and work with sampling mechanisms that fulfill that condition. It would be interesting to understand the relationship between amenability and graph ignorability. It is reasonable to think that the methodology proposed here can be used for dealing with designs that are not amenable. It is our understanding that all the available literature falls into one of two categories: Either they do not discuss ignorability, but assume implicitly that the sampling mechanism of their choice is ignorable (\emph{e.g.},  \citep{Gile}), or they discuss ignorability and then they restrict their discussion to what they regard as ignorable designs  \citep{HandcGile}. In either case, the problem of making inference on a population quantity using a non-ignorable design is not addressed.

An important feature of the methodology we propose is that it is highly modular. By this we mean that the term $p(I \mid \mathcal{G})$ does not have to correspond to RDS. All the arguments hold for any other design that is not ignorable. The choices we made for $p(\alpha)$ and $p(\mathcal{G} \mid \alpha)$ were based on considerations such as simplicity and computational convenience. In principle, nothing prevents the reader from using a different random graph specification or MRF model, as long as Gibbs sampler schemes as the ones proposed at the end of Section \ref{sec-inference} are feasible to implement. 

In recent work, the authors present a set of assumptions that guarantee the asymptotic unbiasedness of the Volz-Heckathorn estimator \citep{SalgaHecka,Hecka2}. These assumptions are:
\begin{enumerate}
\item If individual $i$ can recruit individual $j$ with positive probability, then the probability of $j$ recruiting $i$ must be positive, this is for all $1\leq i<j \leq N$.
\item The graph $\mathcal{G}$ is connected.
\item The proportion of sampled individuals is low enough so assuming that the sampling is with replacement is a reasonable approximation.
\item Respondents are able to accurately report their degree.
\item Each individual selects the subset of peers she (or he) will give the coupons according to an uniform distribution.
\end{enumerate}
Assumptions (a) and (e) are phrased in terms of what is called the \emph{respondent referral function} \citep{BlitzNest}. Note that our approach works for a general $p(I \mid \mathcal{G})$, therefore both assumptions become irrelevant for our framework. Since we can specify $p(\mathcal{G} \mid \alpha)p(\alpha)$ in our approach, we can deal with situations where there is not reasonable to assume $\mathcal{G}$ to be connected (Assumption (b)). Neither for the specification of the model or for the inference we need to assume sampling with replacement as an approximation for $I$, therefore our methodology is blind to Assumption (c). Our method could, in principle, incorporate information regarding the degree of the individuals included in the sample. If we wanted to do so, we could accommodate for a probabilistic model for the reported degree given the real degree. It follows that we could, in principle, deal with situations where Assumption (d) does not hold.

The main sources of uncertainty we consider in our formulation are: (i) the uncertainty due to the underlying network; (ii) the uncertainty due to the sampling mechanism; (iii) the uncertainty due to the fact that we do not know parameters of the model; (iv) the uncertainty induced by the dependence of the responses. One source of uncertainty that could be relevant for this type of application and that we did not discuss is the presence of covariates; these could be incorporated in the random graph model (as in the formulation by \cite{HoffRaftHand}) or in the sampling mechanism via a respondent referral function. This is a direction we plan to pursue in our future work. For the model specifications discussed in this paper, the sampling mechanism is regarded as conditional on the seeds (as in the data analysis) or the seeds are assumed to be sampled uniformly at random (as in the simulation studies). Our methodology could be refined by modelling the distribution of the seeds using distributions different from the uniform; this is an interesting direction for future work. If a probabilistic model for the seeds seems too hard to be formulate explicitly, a sensitivity analysis could be helpful for the practitioner to assess how robust the inferences are on this regard.  The strategy that we propose for obtaining samples for the posterior has $N$ (the maximum number of number of nodes to be augmented to $\mathcal{G}_{\text{INC}}$) as a tuning parameter in our method. This is acknowledged in Section 3.3 and in the data analysis section (Section 5), where we perform a sensitivity analysis with respect to this quantity.

One concern regarding the probabilistic modelling of data gathered via RDS is the feature known as \emph{differential activity}. It is worth to clarify that, while the concept of differential activity for different $Y$ classes (like the concept of homophily) is useful as well as appealing for many practitioners, the framework proposed in this paper is based on the idea that the distribution of the responses (as well as the distribution of the sampling mechanism) should be conditional on the network, rather than modelling the network conditional on the response vector or other features. Obtaining a joint distribution from which we could generate samples that show features that mimic differential activity is possible, given a different specification of the $p(Y \mid \mathcal{G}, \gamma)$ factor, for example, by positing $\mathcal{G}$ comes from a Stochastic Block Model where the (latent) blocks are predictive of the $Y$ classes. Our framework could be extended in that direction.

Our analysis identifies key modelling elements that lead to inferences with good frequentist properties when dealing with data collected through non-ignorable network sampling mechanisms. The proposed estimation strategy achieves higher frequentist coverage with shorter intervals. This is possible because, while the sampling mechanism is non-ignorable, it is possible to write down an accurate model for it, and estimate its parameters form data accurately. These estimates, in turn, provide more information about the inferential target. As argued above, the experimenter who implements a non-ignorable sampling mechanism typically has a good handle on its sources of bias and uncertainty, in theory and practice. We demonstrate the proposed methods via simulations and on a study of the incidence of HIV in Brazil. 

Our research suggests that RDS may not be the best way for collecting information about the parameters of the model, given all the sources of uncertainty. We are able to obtain good point estimates, but the associate uncertainty is higher than what  most practitioners would like to afford. Future work includes taking the data generative process as a starting point (\emph{i.e.}, the terms $p(\mathcal{G} \mid \alpha)$ and $p(Y \mid \mathcal{G},\gamma)$ ) to design sampling mechanisms of the form 
$p(I \mid \mathcal{G})$ that would help to provide better inferences. That would imply establishing criteria for comparing sampling designs on networks, a remarkably unexplored area.

	

%



\bibliographystyle{chicago}
\bibliography{mybib}

\appendix

\section{Annealed Importance Sampling}\label{App:AIS}
The simple importance sampling algorithm (SIS) is an easy to implement algorithm which provides unbiased estimates for partition functions. However, it suffers from the drawback of potentially producing estimates with  infinite variance. It can also perform poorly if the instrumental distribution is very different from the target distribution. 

To overcome these difficulties, \cite{Neal} proposed the annealed importance sampling (AIS) algorithm. This algorithm requires a finite set of functions that  serve as a `bridge' from the instrumental to the target distribution. In this paper, we implemented AIS using the following set of functions
\begin{displaymath}
p_k(\cdot)\propto \tilde{p}(\cdot)^{(1-\beta_k)}p(\cdot)^{\beta_k},\qquad 0\leq k \leq r, 
\end{displaymath}
to set a bridge from the instrumental distribution $\tilde{p}(\cdot)$ to the target $p(\cdot)$. Here $0=\beta_0 < \beta_1 \dots \beta_{r-1} < \beta_r =1$. For this algorithm, it is required that the partition function $Z_0$ of the instrumental distribution is easy to compute. To implement AIS, one also requires a set of transition operators $\left\{ T_k \right\}_{1\leq k \leq r}$ such that, each $T_k$ has $p_k$ as its invariant distribution. These operators take a (putative) Monte Carlo sample from $p_{k-1}$ as  input and produce a sample from a distribution that approximates $p_{k}$ as output.

We now outline the AIS algorithm:
\begin{enumerate}
\item Simulate
         \begin{itemize}
         \item $y_1$ from $p_0(\cdot)$;
         \item $y_2$ from $T_1(\cdot; y_1)$;
         \item \dots
         \item $y_r$ from $T_{r-1}(\cdot; y_{r-1})$.
         \end{itemize}
\item Compute
\begin{displaymath}
w_{\text{AIS}}^{(i)}=\frac{p_1(y_1)}{p_0(y_1)}\times\frac{p_2(y_2)}{p_1(y_2)}\times \dots \frac{p_{r-1}(y_{r-1})}{p_{r-2}(y_{r-1})} \times \frac{p_{r}(y_{r})}{p_{r-1}(y_{r})},
\end{displaymath}
for $1 \leq i \leq M$.
\item Use the values from (b) as weights for an importance sampling algorithm to obtain
\begin{displaymath}
\frac{\hat{Z}_{\text{Target}}}{Z_0}=\frac{1}{M} \sum_{i=1}^M w_{\text{AIS}}^{(i)},
\end{displaymath}
where $\hat{Z}_{\text{Target}}$ is the point estimate for the partition function of the target distribution.
\end{enumerate}
A strong argument for using AIS over SIS, is that, for large values of $M$, the variance of $\hat{Z}_{\text{Target}}$ is proportional to $\frac{1}{rM}$.

\section{More on Simulations}

We fitted an ANCOVA model to assess if the method used to estimate the interval was significant when regarded as a factor. For the case where the model is correctly specified (in terms of random graph model) the other factors to be considered are the size of the network and the density. The response variable is the coverage for the super population parameter $Q_\infty$. Results are summarized in Table \ref{Tab:Ancova1}.

We also fitted an ANCOVA for the case where the model is unspecified (in terms of random graph model) the other factor to be considered  was the density. The response variable is the coverage for the super population parameter $Q_\infty$. Results are summarized in Table  \ref{Tab:Ancova2}.

\begin{table}
\caption{\label{Tab:Ancova1} Results from fitting an ANCOVA model where the coverage of the interval for $Q_\infty$ is the response. Here we assume that the model is well specified. The method used (VH \emph{vs.} Bayesian) appeared as a significant predictor at the $0.05$ level.}
\centering
\fbox{%
\begin{tabular}{| c | c c c c | l |}
 \hline
 Factor & DF & Sum Sq & Mean Sq & F Value & $\Pr(>F)$   \\
\hline
factor(TheMethod)  & 1   &  0.0016000 &  0.00160000  &  8.7549  &  0.011946 * \\
TheSize                  & 1   &  0.0000250 &  0.00002500  &  0.1368  &  0.717934       \\
TheDensity             & 1   &  0.0023569 & 0.00235693   &12.8966  &  0.003705 **   \\
Residuals               & 12  & 0.0021931 & 0.00018276   &                &                       \\
\hline
\end{tabular}}
\end{table}

\begin{table}
\caption{\label{Tab:Ancova2} Results from fitting an ANCOVA model where the coverage of the interval for $Q_\infty$ is the response. Here we assume that the underlying network is either Erdos-Renyi or Small World with re-wiring probability equal to $0.5$. The method used (VH \emph{vs.} Bayesian) appeared as a significant predictor at the $0.05$ level.}
\centering
\fbox{%
\begin{tabular}{| c | c c c c | l |}
 \hline
 Factor & DF & Sum Sq & Mean Sq & F Value & $\Pr(>F)$   \\
\hline
factor(TheMethod)  &  1   &  0.0025000   &  0.0025000  &  7.1104   &  0.020542 *  \\
factor(TheRG)         &  1   &  0.0042250   &  0.0042250  & 12.0165 &  0.004661 ** \\
TheDensity              &  1   &  0.0004308   &  0.0004308  & 1.2253   &  0.290018     \\ 
Residuals                & 12   & 0.0042192   &  0.0003516   &               &                     \\
\hline
\end{tabular}}
\end{table}

\end{document}